\documentclass[a4paper,usenatbib]{mnras}

\usepackage[T1]{fontenc}
\usepackage{ae,aecompl}

\usepackage{amssymb}	% Extra maths symbols
\usepackage{natbib}
\usepackage{graphicx}
\usepackage{rotating,times,graphicx,latexsym}
\usepackage{color}
\usepackage{longtable}
\usepackage{amsmath}
\usepackage{lscape}
\usepackage{lipsum} % To generate text
\usepackage{array}
\usepackage{soul} % To add highlighting of text
\usepackage{xargs} % Use more than one optional parameter in a new commands
\usepackage[pdftex,dvipsnames]{xcolor}  % Coloured text etc.
\usepackage{amsmath}
\usepackage[colorinlistoftodos]{todonotes}
\newcommandx{\tobedone}[2][1=]{\todo[linecolor=red,backgroundcolor=red!25,bordercolor=red,inline,#1]{#2}}
\newcommandx{\changed}[2][1=]{\todo[linecolor=blue,backgroundcolor=blue!25,bordercolor=blue,inline,#1]{#2}\noindent}
\newcommandx{\thiswillnotshow}[2][1=]{\todo[disable,#1]{#2}}
\newcommandx{\ak}[2][1=]{\todo[linecolor=Orchid,backgroundcolor=Orchid!25,bordercolor=Orchid,inline,#1]{#2}\noindent}
\newcommandx{\df}[2][1=]{\todo[linecolor=BurntOrange,backgroundcolor=BurntOrange!25,bordercolor=BurntOrange,inline,#1]{#2}\noindent}
\newcommandx{\jcw}[2][1=]{\todo[linecolor=SpringGreen,backgroundcolor=SpringGreen!40,bordercolor=SpringGreen,inline,#1]{#2}\noindent}
\newcommand{\mic}{\,$\mu$m~}

\newcommand{\dg}{$^{\circ}$}

\newcommand{\Msolar}{\,M$_{\odot}$~}

\newcommand{\ha}{H$\alpha$~}

\newcommand{\hii}{H{\sc ii}~}

\title[Shape Analysis of \hii Regions]{Shape Analysis of \hii Regions -- I. Statistical Clustering}

\author[J. Campbell-White et al.]{
Justyn Campbell-White,$^{1}$\thanks{E-mail: jc849@kent.ac.uk}
Dirk Froebrich,$^{1}$
Alfred Kume$^{2}$
\\
$^{1}$Centre for Astrophysics and Planetary Science, The University of Kent, Canterbury, CT2 7NH\\
$^{2}$School of Mathematics, Statistics and Actuarial Sciences, The University of Kent, Canterbury, CT2 7FS\\
}

\date{Accepted XXX. Received YYY; in original form ZZZ}

\pubyear{2018}

\begin{document}
\label{firstpage}
\pagerange{\pageref{firstpage}--\pageref{lastpage}}
\maketitle

\begin{abstract}
We present here our shape analysis method for a sample of 76 Galactic \hii regions from MAGPIS 1.4\,GHz data. The main goal is to determine whether physical properties and initial conditions of massive star cluster formation is linked to the shape of the regions. We outline a systematic procedure for extracting region shapes and perform hierarchical clustering on the shape data. We identified six groups that categorise \hii regions by common morphologies. We confirmed the validity of these groupings by bootstrap re-sampling and the ordinance technique multidimensional scaling. We then investigated associations between physical parameters and the assigned groups. Location is mostly independent of group, with a small preference for regions of similar longitudes to share common morphologies. The shapes are homogeneously distributed across Galactocentric distance and latitude. One group contains regions that are all younger than 0.5\,Myr and ionised by low- to intermediate-mass sources. Those in another group are all driven by intermediate- to high-mass sources. One group was distinctly separated from the other five and contained regions at the surface brightness detection limit for the survey. We find that our hierarchical procedure is most sensitive to the spatial sampling resolution used, which is determined for each region from its distance. We discuss how these errors can be further quantified and reduced in future work by utilising synthetic observations from numerical simulations of \hii regions. We also outline how this shape analysis has further applications to other diffuse astronomical objects.

\end{abstract}


%%%%%%%%%%%%%%%%%%%%%%%%%%%%%%%%%%%%%%%%%%%%%%%%%%

%%%%%%%%%%%%%%%%% BODY OF PAPER %%%%%%%%%%%%%%%%%%

\section{Introduction}
\label{sec:intro}
Massive stars capable of exciting an \hii region have lifetimes up to $\sim$\,10 million years. In this time, a significant amount of feedback is delivered to the surrounding interstellar medium (ISM) via the \hii region, outflows, stellar winds, and ultimately supernovae explosions. The ISM also influences the morphologies of \hii regions due to its non-homogeneous nature. \hii regions form around their ionising source(s) within $10^5$\,years, sweeping up material from the ISM into a dense shell surrounding hot dust and ionised gas \citep{1977ApJ...218..377W}. Continuum emission from the ionised gas is readily observed at cm wavelengths, along with coincident mid-infrared (MIR) emission distinguishing these objects as thermal sources. The dense shells surrounding \hii regions are well traced by strong bands of excited polycyclic aromatic hydrocarbons (PAHs), for example the 7.7\mic and 8.6\mic bands \citep{2008ApJ...681.1341W} highlight the shells in the \textit{Spitzer}-GLIMPSE survey \citep{2003PASP..115..953B}. \citet{2010A&A...523A...6D} found that the ratio of surface brightness of 24\mic emission (from the MIPGASL survey, which traces the hot dust \citep{2009PASP..121...76C}) to 20\,cm emission peaks at the centre of the \hii region, but is less spatially coincident with increasing distance to the centre. %This is due to the differing temperatures of the dust, which is higher near the ionising source, suggesting that the dust grains are heated by Lyman continuum photons from the ionising source, rather than by the Lyman alpha photons which are abundant throughout the ionised region.  

In this work, we look to analyse the high resolution image data of \hii regions using detailed shape analysis. Thus far, radio continuum images of \hii regions have been observed to share a common morphology of a closed disk, which is inferred as a projection of the extended Str\"{o}mgren sphere around an ionising source(s) (see e.g \citet{1967ApJ...147..471M,1982A&A...112....1B,2000Ap&SS.272..169F,2006ApJS..165..338Q}). Similarly, MIR bubbles (the larger of which are all confirmed to be the result of \hii regions \citep{2011ApJS..194...32A,2010A&A...523A...6D}) have been characterised by their common morphologies of ring like structures, and studies have catalogued their radii, eccentricities and shell thickness's (e.g. \citet{2006ApJ...649..759C,2007ApJ...670..428C,2010ApJ...709..791B,2011ApJ...742..105A,2012MNRAS.424.2442S}). In each of these cases, inhomogeneities in the natal molecular clouds can lead to perturbations from these ideal morphologies. If a link can be found between the shape of the \hii regions and the host physical conditions, we can use this to better understand the affinity between massive stars, their formation and the surrounding ISM. 

In terms of classifying \hii regions and bubbles via their observed shape, a broad morphological scheme was introduced by \citet[hereafter C06]{2006ApJ...649..759C,2007ApJ...670..428C}, who first studied stellar bubble images from the GLIMPSE data. They classified almost 600 MIR bubbles along the Galactic plane by the shape of their shells, e.g. closed, broken, group. This approach was also used by \citet{2011ApJS..194...32A} and \citet{2012ApJ...759...96B} for the Green Bank and Arecibo \hii Region Discovery Survey catalogues, respectively. Despite the fact that only half of all \hii regions are observed as MIR bubbles \citep{2011ApJS..194...32A}, it is easier to distinguish features by eye in the MIR shells than the radio continuum disks. More recently, \citet{2014ApJS..212....1A} categorised over 8000 regions in the WISE Catalog of Galactic HII Regions based on the presence or absence of radio emission and do not include any morphological flags.

\hii regions are known to trace the locations of active star formation (SF) (e.g. \citet{2012MNRAS.421..408T,2012ApJ...755...71K}), and due to the clustered nature of massive SF, insight into the physical properties of \hii regions allow for inferences to be made as to the initial conditions of massive cluster formation. Difficulties in determining accurate distances to Galactic \hii regions lead to many statistical investigations having low completeness. In this work, we take advantage of the homogeneity of the high resolution radio images available of \hii regions, to systematically compare their shapes. Shape provides an unbiased, intrinsic characteristic of an object, which can be readily compared through statistical means. In this paper, we present our shape analysis method for the unsupervised statistical clustering of \hii regions. In a subsequent paper (Campbell-White et al. in prep.),  we will compare the results from our observed sample to those from synthetic observations of numerical simulations of \hii regions. The aim of this work is to find associations between the region shapes and physical parameters/initial conditions of cluster formation to produce a training set for supervised classification of \hii regions.

This work is organised as follows: In Section\,\ref{sec:data}, we describe the source and extraction of the observational data. In Section\,\ref{sec:method_results}, we provide details of the statistical methodology and present our results. In Section\,\ref{sec:discussion}, we discuss the physical properties of the \hii regions with respect to our findings; discuss outliers in the data, and propose future applications of our method. This is all summarised and concluded in Section\,\ref{sec:conclusions}.

\section{\hii Region Image Data}
\label{sec:data}

The WISE Catalogue of Galactic \hii regions \citep{2014ApJS..212....1A} is currently the largest compiled catalogue of \hii regions. It comprises details of 8399 identified and candidate \hii regions, of which 1524 are categorised as \textquoteleft Known', with observed \ha spectroscopic and/or Radio-Recombination-Line (RRL) emission. The remaining regions are flagged as groups, candidates and radio-quiet regions identified by their MIR morphology in the WISE images.

High resolution MIR images of stellar bubbles from the \textit{Spitzer}-GLIMPSE survey were first studied in C06. In that work, they obtained estimates for the angular radii, eccentricity, position angle and thickness of almost 600 bubbles. This was from visual inspection of the MIR images by the authors. The number of identified bubbles from the \textit{Spitzer} images was then increased by an order of magnitude by the Milky Way Project DR1 \citep[MWP,][]{2012MNRAS.424.2442S}. The MWP bubble catalogue comprises details of 5106 stellar bubbles, identified by citizen scientists\footnote{Using the Zooniverse platform: https://www.zooniverse.org/} and is split into two sub-catalogues of 3744 large and 1362 small bubbles. The small bubbles are those which resemble the required MIR morphology of a stellar bubble, but are too small for the platform's ellipse drawing tool. Properties of the large bubbles were obtained, all reduced from the compiled crowd-sourced data, corresponding to those of C06.

Although well categorised, the MIR data of the stellar bubbles are subject to many inhomogeneities (such as point sources and filamentary structures) that could prove difficult as a starting point for the shape analysis we utilise in this work. Instead, we chose to start with the high resolution 20cm radio continuum images from the Multi-Array Galactic Plane Imaging Survey \citep[MAGPIS,][]{2006AJ....131.2525H}. MAGPIS combined VLA images with those from the 1.4 GHz 100m Effelsberg telescope to correct for missing fluxes in the extended emission regions. The resulting MAGPIS images have an average angular resolution of 5.8", with a pixel scale of 2" and 1$\sigma$ sensitivity of $<0.15$\,mJy. MAGPIS covers $|b| < 0.8$\dg and $48.5$\dg $> l > 5$\dg. Within this range, there are 710 \textquoteleft Known' WISE \hii regions, 405 of which have a determined distance in the WISE catalogue. With the intent of analysing the MIR morphology of the regions in future work, we identified which of these 405 \hii regions were positionally coincident with at least one MWP large bubble, which resulted in 243 regions. 

\begin{figure*}
	\includegraphics[width=\textwidth]{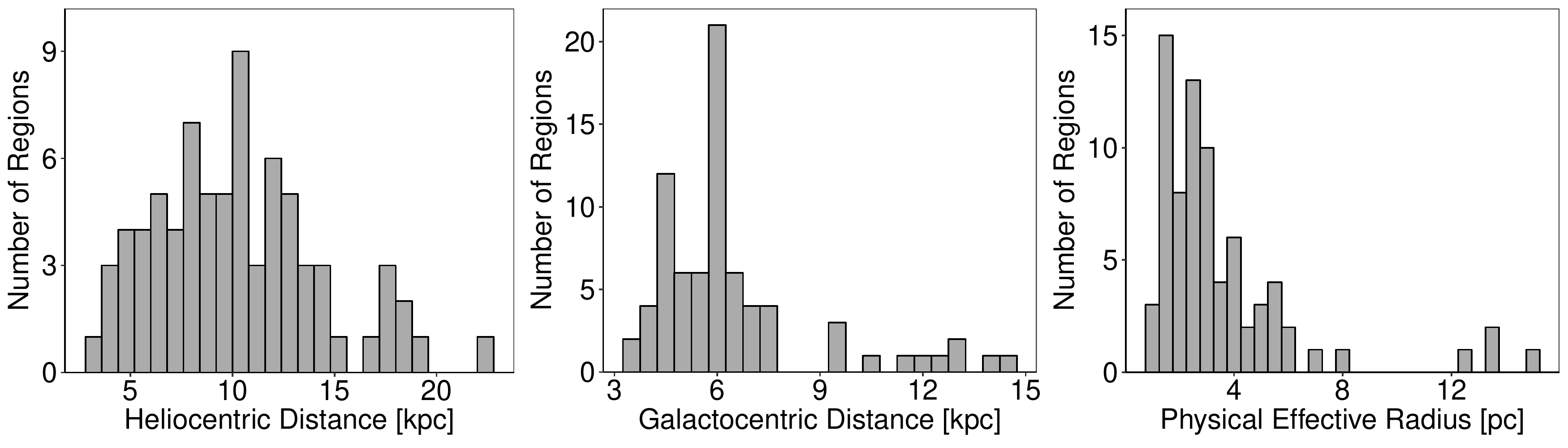}
	\caption{Distributions of distances of our sample of 76 \hii region distances (left), Galactocentric distances (middle) and physical effective radii (right). Distances were taken from \citet{2014ApJS..212....1A}, and used to determine the other two parameters. The effective radius was taken as the mean of the semi-major and -minor axes of the contoured regions.}
	\label{fig:gcd_dist_reff}
\end{figure*}

The morphologies of this sample of \hii regions at a given surface brightness were extracted using image contouring. To generate contours from each region, the signal value was first determined by applying sigma clipping with five iterations to each image tile. Contour levels were then applied with values of 0.5, 1, 1.5 and 2$\sigma$ above the mean, with a smoothing of 3 pixels, which accounted for the 5.8" beam size of MAGPIS, given the 2" pixel scale. This enabled us to determine a systematic boundary of each region, whilst accounting for the varying fluxes and noise. Visual inspection of the contouring procedure showed that the 1$\sigma$ contour was most effectively capturing the \textquoteleft edge' of the \hii regions in the MAGPIS data. Both the 1.5 and 2$\sigma$ levels frequently missed some of the extended emission features and were concentrated around the brightest pixels. The 0.5$\sigma$ level extended out into the image noise for many of the regions. Hence, we selected the 1$\sigma$ contour levels to describe the shape of each \hii region. Further testing of how different sigma levels affect the methodology and results are outlined in Sect.\,\ref{sec:group_valid}. 
	
We selected our final sample of \hii regions from the 1$\sigma$ contoured image tiles. To test our shape analysis method, we were more stringent with our selection than if we were simply intending to classify all of the \hii regions within our larger data set. Regions that were at the edge of image tiles were disregarded, as were ones where the contouring procedure did not generate at least one closed boundary around the region. We thus arrived at a sample of $n=76$ \hii regions, comprising different shapes and sizes. The contouring procedure produced a set of $l,b$ coordinates for the boundary of each region. After translating the centre of each set to the origin of the coordinate axes, each region's distance was then used to change from angular Galactic degrees to spatial parsec coordinates. Since all regions were within $\pm0.8 ^\circ$ of the Galactic Plane, the correction factor to account for the spherical coordinate system can be neglected.

Distances to the \hii regions were taken from the WISE catalogue. The most common technique for obtaining these distances is kinematically via a measured line of sight radial (LSR) velocity. \citet{2012ApJ...754...62A} found that the errors associated to kinematic distance estimates are between 10\% and 20\% for the entire Galactic Plane, due to uncertainties in rotation curves and the Solar circular rotation parameters. There are also further systematic errors for kinematic distances of objects tangential to our line of sight, and those located within the Solar circle are subject to the kinematic distance ambiguity, where each LSR velocity value has a near and a far associated distance. The average error in the distances to \hii regions in the first quadrant of the Galaxy from the WISE catalogue is 15\% \citep{2014ApJS..212....1A}. The distribution of Heliocentric distances of our sample of \hii regions range from 3.5 to 22.6\,kpc with a peak of $\sim$\,11\,kpc (Fig.~\ref{fig:gcd_dist_reff}, left).
Galactocentric distances are shown in Fig.~\ref{fig:gcd_dist_reff}, middle with a peak at $
\sim$\,6\,kpc. The mean physical effective radius of each region was determined from the contoured boundary and most are less than 5\,pc (Fig.~\ref{fig:gcd_dist_reff}, right).

Five of the \hii regions in our sample are positionally coincident with a young (< 10\,Myr) star cluster from the Milky Way Global Survey of Star Clusters catalogue \citep[MWSC,][]{2012A&A...543A.156K} (G022.761-00.492 \& BDSB\_117, G034.256+00.136 \& BDSB\_127, G042.103-00.623 \& BDSB\_131, G045.475+00.130 \& BDSB\_136, and G045.825-00.291 \& BDSB\_137). Each of the clusters from MWSC are flagged as \textquoteleft nebulous', hence they may be physically associated with the \hii regions. Three of the clusters, BDSB\_136, BDSB\_137 and BDSB\_127 have distance estimates 22\% larger than the corresponding \hii region distance from the WISE catalogue. The distances to BDSB\_131 and BDSB\_117 are 43\% higher and 36\% lower, respectively. Due to the embedded nature of massive SF, it is expected that few identified star clusters would be positionally coincident with \hii regions. Since these examples are just coincident, and not shown to be associations, we instead keep the distance estimates from the WISE catalogue.

\section{Statistical Methods \& Results}
\label{sec:method_results}

We present here our statistical shape analysis using the sample of 76 selected \hii regions. The \hii regions/bubbles in our data share a common morphology of a rounded extended disk for the former, and ring like structure for the latter. It is this property that has allowed them to be readily identified from visual inspection of the radio continuum or mid-IR images. As outlined in C06, however, few bubbles display a well defined circular symmetry, with many highly elliptical or only partial rings. This was also found by \citet{2012MNRAS.424.2442S} in the MWP data set. Recent focused studies on individual bubbles have looked at perturbations of shells and association with massive clumps as signs of triggered star formation (e.g. \citet{2012A&A...544A..39J,2014A&A...569A..36X,2017ApJ...834...22D}). However, the availability of high angular resolution wide field images, that reveal the fine detail of \hii region boundaries at sub-pc level, allows for robust shape analysis techniques to be carried out across a larger sample. 

An object's shape is defined as \textquoteleft all the geometric information that remains when location, scale and rotational effects are filtered out' \citep{10.2307/1426091}. If one would like to preserve information as to the object's scale, size-and-shape is considered. Two objects have an equal size-and-shape if a rigid body transformation of one can match it exactly to the other. Shape is hence used as an invariant measure that provides an intrinsic property of objects under investigation. A caveat when dealing with astrophysical objects, however, is that one must ensure that the extraction of shape for the object in question is systematic and repeatable. This is because changing the threshold values for image data would inevitably lead to deviations in the identified boundaries. An object's shape is quantified along its boundary by a finite number of landmarks. A shape descriptor conveys this information, depending on how the landmarks are interpreted. For example, the effective radius measurements used for the MWP and C06 bubbles is the shape descriptor for the average of the semi major and minor axes of the ellipses fitted to the bubbles.

\subsection{Curvature Distribution}
The shape descriptor we use in our analysis is the curvature, $k$, which is defined as the reciprocal of the radius, hence this descriptor preserves the object's size-and-shape invariance. Landmarks were taken as the coordinates of the boundary of each shape, having been transformed from an angular to a spatial scale (as outlined in Sect.~\ref{sec:data}). This boundary was taken as $f(t)=x(t)y(t)$, where $t$ is the spatial interval between each ($x,y$) Cartesian coordinate pair. For each object, the curvature was calculated at multiple points along the boundary, $f(t)$, by the following equation, where the primes denote the first and second derivatives with respect to $t$. 

\begin{equation}
	k(t)=\frac{\vert x^{\prime} y^{\prime\prime} - x^{\prime\prime} y^{\prime}\vert} {(x^{\prime2} + y^{\prime2})^{\frac{3}{2}} }
	\label{eq:curvature}
\end{equation}

\begin{figure}
	\includegraphics[width=\columnwidth]{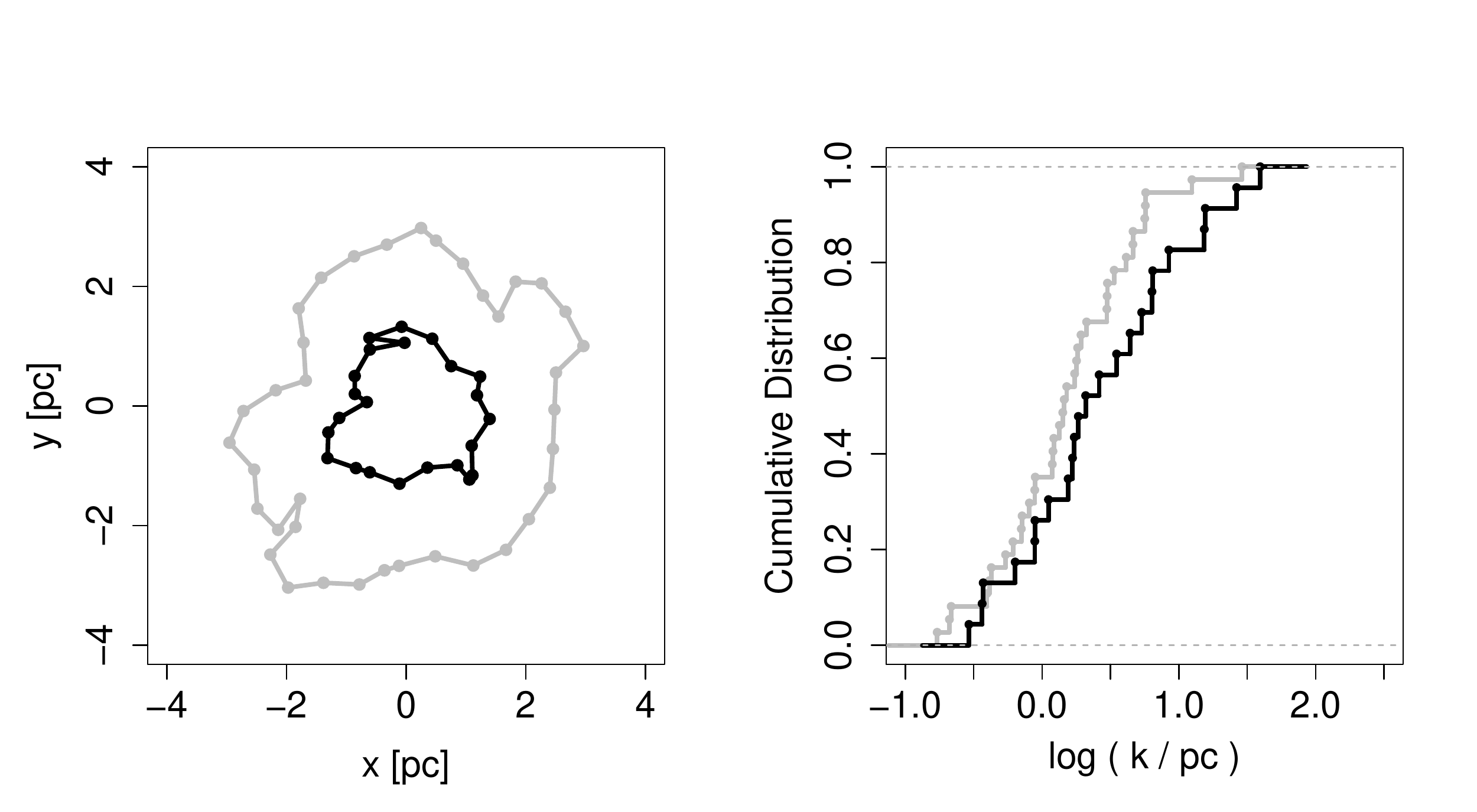}
	\caption{Left: Boundaries of \hii regions G038.840$+$00.495 (grey) and G018.152$+$00.090 (black) with centred, spatial coordinates. Points indicate interpolation spline \textquoteleft knots', where the curvature $k$ was calculated. Right: Corresponding empirical cumulative distribution functions for the curvature values along the boundary of each \hii region.}
	\label{fig:rplot_ecdf}
\end{figure}

These derivatives were obtained by fitting cubic interpolation splines to $f(t)$. The points for interpolation (spline knots) could then be set at a given spatial resolution, whereby the derivatives; and thus the curvature, were determined at each knot. This allowed for each region's shape to be represented by its corresponding curvature distribution, $k(t)$. The mean knot interval for the entire data was set to 0.54\,pc, which corresponds to the spatial beam size of the MAGPIS survey at a distance of 19.2\,kpc - the lower error bound of the heliocentric distance to the furthest object in the sample. This limiting sampling resolution aimed to ensure that no bias was given to objects at a nearer distance, where the spatial resolution of the images is higher. For such objects where the initial number of data points in $f(t)$ was far greater than the number of spline knots, the boundaries were notably under-sampled, leading to the small amount of variation in the 0.54\,pc knot intervals. The resulting knot intervals were normally distributed with $\mu$ = 0.54\,pc and $\sigma$ = 0.16\,pc. Resulting affects from changing this knot interval depend upon whether or not the overall description of the object's shape changes and are discussed in more detail in Sect.\,\ref{sec:group_valid}. Figure \ref{fig:rplot_ecdf} shows an example of two different shape boundaries with centred, spatial coordinates (\textit{left}), with the corresponding empirical cumulative distribution functions (CDF) for the curvature at each knot (\textit{right}). The use of CDFs for summarising the curvature distributions allows for an unbiased, consistent estimator of the population, with no loss of information due to binning of data.

\subsection{Anderson-Darling Test}

In order to group objects based on their curvature distributions, the two-sided Anderson-Darling (A-D) test statistic \citep{anderson1952} was used. 
The A-D test is a statistical hypothesis test, where the null hypothesis when applied to two samples is that they are subsets of a single parent population. The two-sample A-D test, $T_{AD}$ \citep{pettitt1976two}, generalises to the following formula:

\begin{equation}
T_{AD}=\frac{1}{nm} \sum_{i=1}^{n+m}\frac{(N_iZ_{(n+m-ni)})^2}{iZ_{(n+m-i)}}
\label{eq:a-d}
\end{equation}

\noindent where $Z_{(n+m)}$ are the combined and ordered samples $X_{(n)}$ and $Y_{(m)}$ (for respective sample sizes $n$ and $m$), and $N_i$ are the number of observations in $X_{(n)}$ that are less than or equal to the $i^{\rm th}$ observation in $Z_{(n+m)}$. The test statistic $T_{AD}$ represents a dissimilarity measure between the two samples, whereby the null hypothesis is rejected for large $T_{AD}$, and $T_{AD}=0$ for identical distributions.

The A-D test is similar to the Kolmogorov-Smirnoff (K-S) test, such that it compares two CDFs. Such statistical tests are useful as they do not require binning of continuous data, hence no information is lost by arbitrary selection choices. 
The K-S test is most suitable when the CDFs differ globally, since it only computes the maximum difference between the two CDFs. The K-S test is also insensitive when differences in the CDFs are most prominent at the tails of the distributions, due to the convergence of the CDF at 0.0 and 1.0. The A-D test, however, measures the sum of square deviations between each CDF, accounting for multiple discrepancies, and is appropriately weighted towards the tails of the distribution, thus addressing both of these issues. For further information of comparisons between the A-D test and the K-S test see e.g. \citet{2006ASPC..351..127B}, \citet{2009ApJ...702.1199H} and \citet{engmann2011comparing}.

The two-sided A-D test statistic was computed pair-wise for each \hii region using the curvature CDFs. This resulted in a symmetrical $N\times N$ matrix of $T_{AD}$ dissimilarity measures. 
 
\begin{figure*}
	\includegraphics[width=\textwidth]{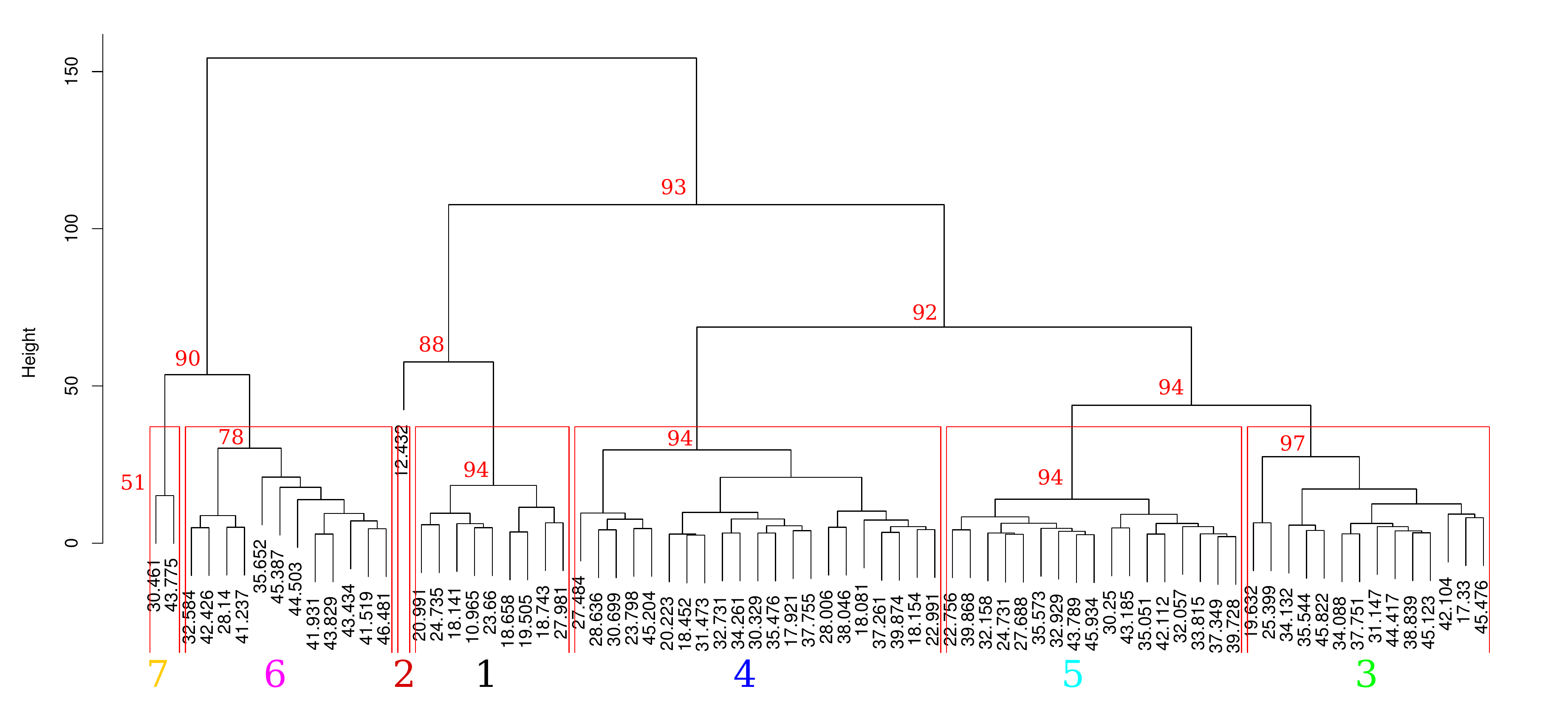}
	\caption{Dendrogram showing the results of our hierarchical cluster analysis of the shapes of our sample of \hii regions. Each \hii region is placed at the bottom of the dendrogram, with groups shown as merges on the dendrogram. The height score of a group is taken from Ward's agglomerative method. Red boxes delineate groups obtained by cutting the dendrogram at a constant height, which are indexed 1 -- 7 at the bottom of the figure. Numbers above merged groups represent the $p$-values (in \%) obtained from multi-scale bootstrap re-sampling of the data. }
	\label{fig:dendrogram}
\end{figure*}

\subsection{Hierarchical Cluster Analysis}
To avoid ambiguity between the terms \textquoteleft star cluster' and \textquoteleft statistical cluster', we will henceforth refer to the statistical cluster groups produced from hierarchical clustering as \textquoteleft groups'. 

The hierarchical clustering used in this analysis requires a distance matrix as the input. The symmetrical matrix of dissimilarity measures from the pair-wise A-D tests was hence converted to a distance matrix using the Euclidean transformation: 
%\df{If you refer to equations then it is either Fig.~uation\,1, when you start a sentence with it, or it is Eq.\,1 when it is not at the start of a sentence.}

\begin{equation}
D_{xy}=\sqrt{\sum_{i=1}^{N}\left(x_i - y_i \right)^2}
\label{eq:dist}
\end{equation}

\noindent  where $x$ and $y$ are the vectors of dissimilarities between the respective regions. For example, $D_{14}$ is the Euclidean distance between regions 1 and 4, which accounts for the dissimilarities of each region with respect to all other regions. This allows for groups to be more accurately determined as outlier distances are highlighted in Euclidean space. The resulting distance matrix is again symmetric, with $D_{x=y}=0$. 

Hierarchical cluster analysis was performed on the distance matrix, with the method \textquotedblleft ward.D2\textquotedblright\footnote{The \textquotedblleft ward.D\textquotedblright~ method yields the same clustering results when using $D_{xy}^2$ as the input}, which implements the \citet{ward1963hierarchical} clustering criterion \citep{murtagh2014ward}. \citet{ferreira2009comparison} found that Ward's method outperformed other hierarchical clustering methods (such as average-, complete- and single- linkage), especially when there were no large differences among group sizes. Ward's agglomerative method is based on a sum-of-squared differences criterion for producing groups. It ensures that at each hierarchical step, the distances of the newly created group to the rest of the data is selected such that the within-group-dispersion is minimised (as explained below, Eq.\,\ref{eq:lw}). This dispersion is proportional to the squared Euclidean distance between group centres, rather than, for example, taking the maximum distance of an object in each group, as adopted in the complete-linkage method.

To iteratively form the groups using Ward's method, each value in the input distance matrix, $D$,  ($i,j,k,~etc. \subset D$, which in our case are the Euclidean distances between regions based on their pair-wise A-D test statistic) starts as a singleton group ($C_i, C_j, C_k, etc.$). The closest pair of singletons are then agglomerated, into a new group. At each iteration, as the groups are formed and grow, inter-group distances, $d$, are then redefined from the new group centres (of group sizes $n$), following the \citet{lange1967general} dissimilarity update formula:

\begin{equation}\label{eq:lw}
	\begin{split}
	d^2(C_i\cup C_j, C_k)=\frac{n_i+n_k}{n_i+n_j+n_k}d^2(C_i,C_k)+\\
	\frac{n_j+n_k}{n_i+n_j+n_k}d^2(C_j,C_k)-\frac{n_k}{n_i+n_j+n_k}d^2(C_i,C_j)
	\end{split}
\end{equation}

\noindent  At each iteration, for $N$ groups (where $N$ originally equals the number of objects in the input distance matrix, and then decreases after each iteration), there are $N-1$ agglomerations. The process is repeated until the final two groups are joined. 

The resulting hierarchical structure is plotted on a dendrogram (Fig.~\ref{fig:dendrogram}), which shows each individual object, termed a \textquoteleft leaf' and the \textquoteleft branches' that join them. The height is the distance value at which the leafs/branches merge. For pairs of leaves, this is the distance value from the input Euclidean distance matrix, $D$. For all merges higher than this, the height value is obtained from Ward's method using Eq.~\ref{eq:lw}. Rather than considering each leaf and how they form branches, one can instead consider the first split in the dendrogram to be the two largest distinct groups, with more groups forming at each subsequent split as one looks further down the dendrogram. In Fig.\,\ref{fig:dendrogram}, the delineated boxes in red represent seven groups obtained by cutting the dendrogram at a constant height. The numbering of the groups is arbitrary, assigned by the algorithm following the sequence in which it considers each object. The position of groups in the dendrogram, however, is not arbitrary, with tighter groups (with a lower height score) positioned starting from the left at merges. By looking top-down, we see that groups 7 \& 6 are separated from the rest of the groups at the highest level. We should hence expect to find most difference between these sets of regions. Considering the right side of this first split, we note that group 1 and the outlier object in group 2 are separate from groups 4, 5 \& 3. The following subsection discusses these groups in more detail.

\subsection{Groups from Clustering}

\subsubsection{Visual inspection and CDFs of groups}
A selection of the \hii regions sorted into each of the groups from the Fig.\,\ref{fig:dendrogram} dendrogram can be seen in Appendix\,\ref{app:group_images}\,\footnote{Available in the online version of this paper.}. It is clear from the images that the amount of local curvature variance along the region boundaries changes between each group. Those grouped on the left hand side of the dendrogram (labelled groups 6 \& 7 in Fig.\,\ref{fig:dendrogram}) possess the highest amount of local curvature variation along their boundaries. We also note visual similarities between groups originating from common parent branches - e.g. groups 5 \& 3. Group 1 appears to host the most unperturbed regions, with low amounts of local curvature variance. The regions in group 4 display a range of morphological features, including boundaries that are circular, elongated, and largely perturbed. This group may hence represent the \textquoteleft average' regions, with no explicitly defining features which would place them in to one of the other groups. It is not immediately apparent why the outlier region in group 2 is placed into a single group. A possible explanation is that the amount of interpolation points used to determine this region's curvature distribution is far greater than most regions, since it is located at the furthest distance in the sample. We discuss this region in more detail in Section\,\ref{sec:outliers}.  

\begin{figure}
	\includegraphics[width=\columnwidth]{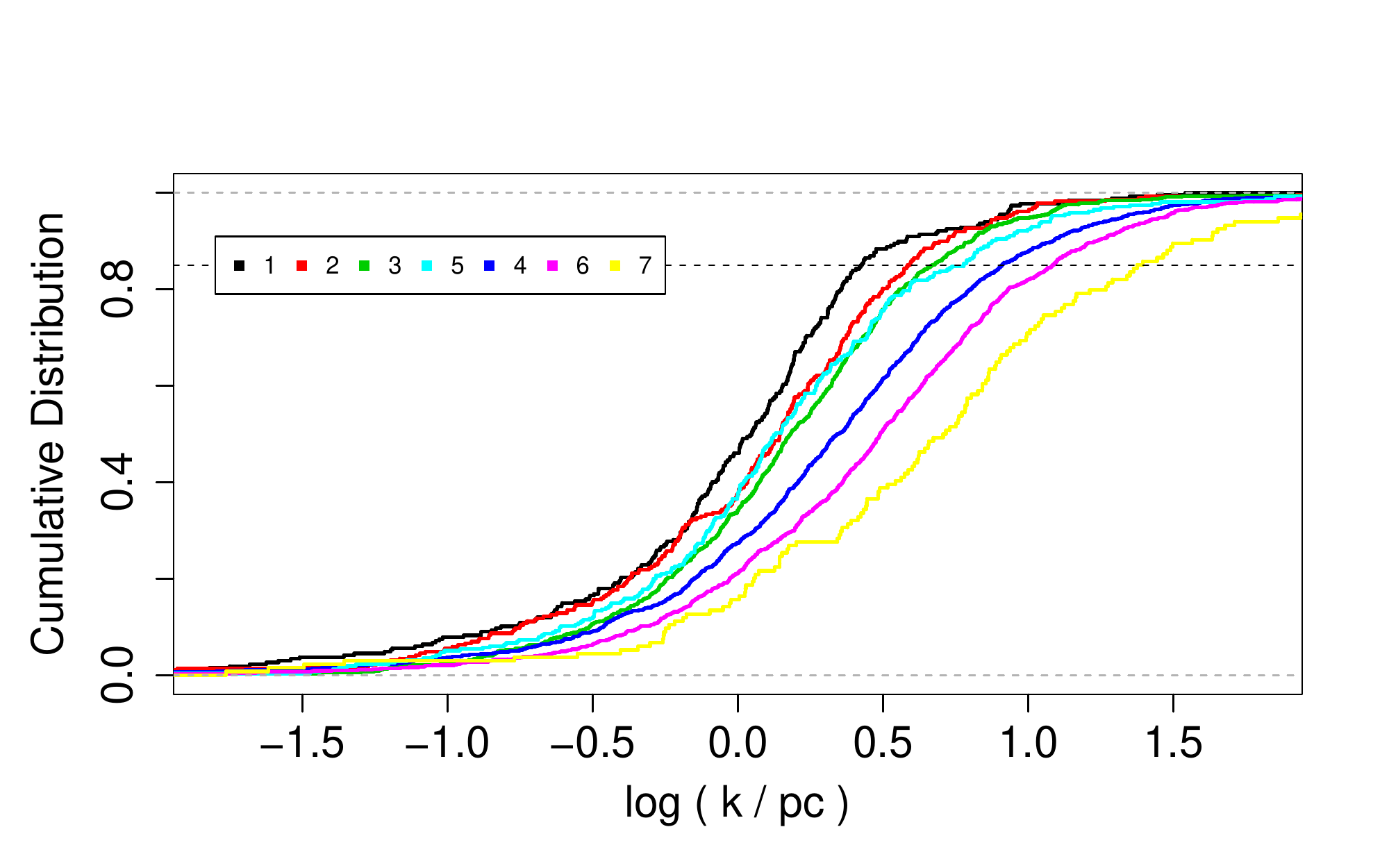}
	\caption{Cumulative distribution functions for the curvature values of all \hii regions belonging to the seven groups identified in Fig. \ref{fig:dendrogram}. Note that this figure has been truncated at log ($k$/pc) = $\pm$1.8 to highlight the differences between CDFs at the centre of the distributions. The maximum curvature values for group 7 are significantly higher than the other groups.}
	\label{fig:clust_cdf}
\end{figure}

For some of the regions, it is hard to tell visually why they are placed into a given group. In order to more quantitatively determine how the clustering procedure is grouping the regions, we considered the CDF of all $k$ values within each of the seven groups identified in the dendrogram. The resulting plot is shown in Fig.~\ref{fig:clust_cdf}. It has been truncated to highlight the global differences between each CDF. The pair of objects in group 7 have the largest fraction of high $k$ values, followed by the objects in group 6. These two groups merge with the rest of the groups at the highest point on the dendrogram, however, group 4 which displays the next largest amount of high $k$ values, is in the middle of the groups on the right side of the first split. This suggests that it is not only the amount of high $k$ values that is delineating the groups. The CDFs of groups 3 and 5 cross multiple times and if the cut on the dendrogram was made higher, these two groups would be the first to merge. Group 1 appears to have the lowest $k$ values, with the outlier region in group 2 having a distribution that lies between group 1 and groups 3 \& 5. Since it was the CDFs that were considered in the pair-wise A-D tests, it is expected that the combined CDFs for each group would show distinct differences. We can further visualise how the hierarchical clustering has generated groups by utilising an ordination technique for the input distance matrix.

\subsubsection{Multi-Dimensional Scaling}

\begin{figure}
	\includegraphics[width=\columnwidth]{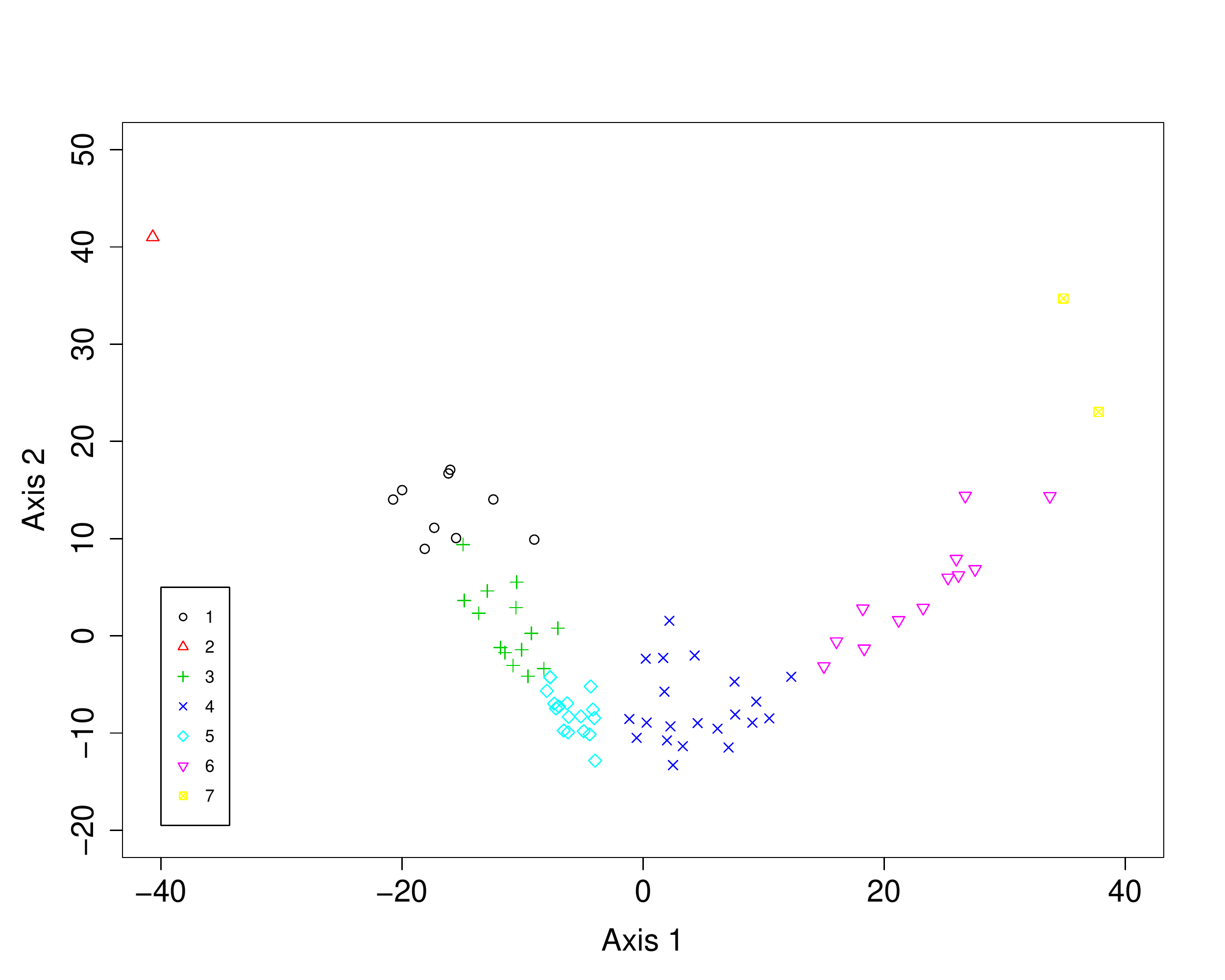}
	\includegraphics[width=\columnwidth]{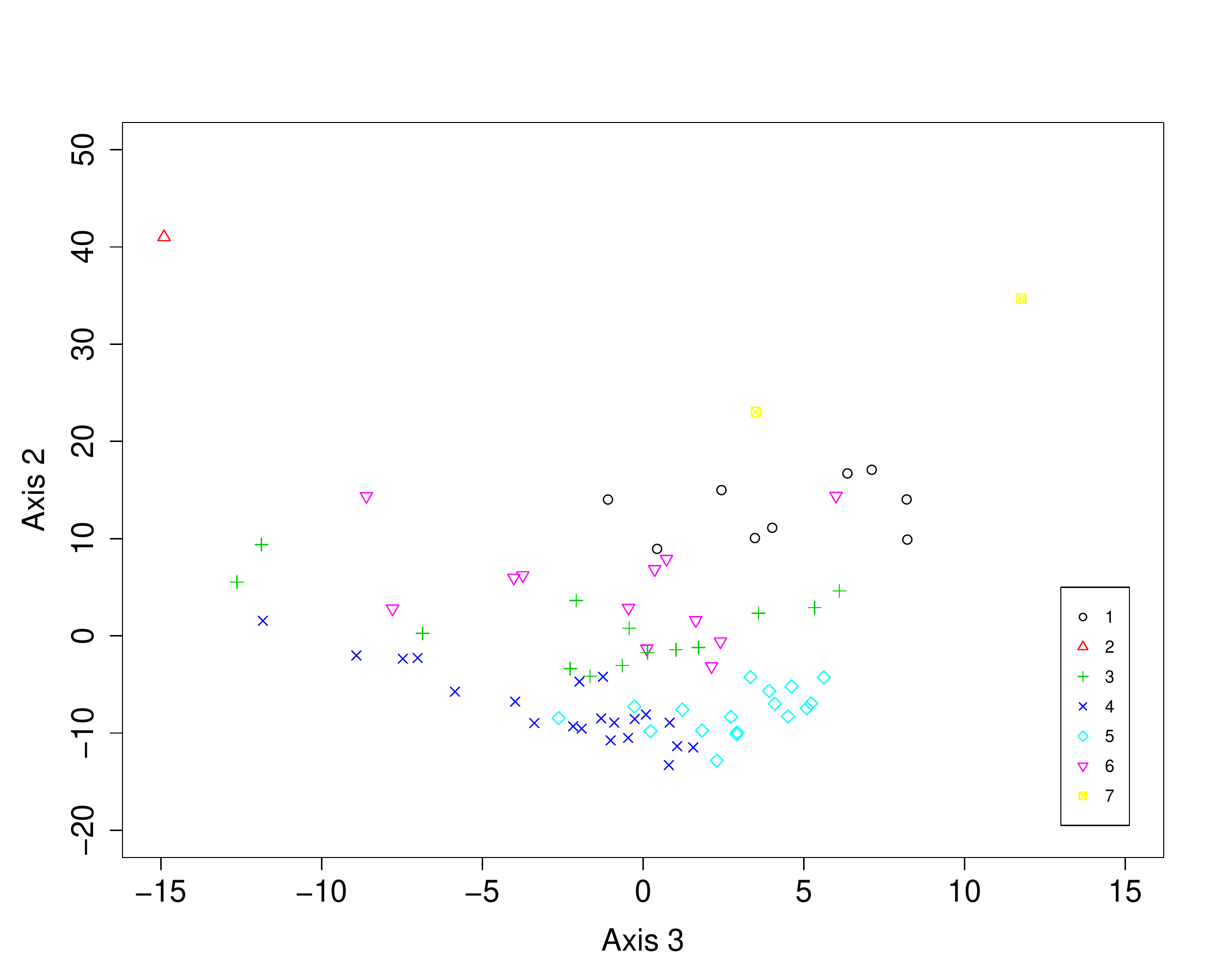}
	\caption{Ordination graphs showing the results of Multi-Dimensional Scaling of the distance matrix used for the hierarchical clustering. Each point represents one \hii region, with colours and symbols matching the groups identified on the dendrogram in Fig.\,\ref{fig:dendrogram}. The first and second principal coordinate axes are shown in the upper panel and the third and second in the lower. Together, these three axes represent 97\% of the variability in the distance matrix.}
	\label{fig:mds}
\end{figure}

The input distance matrix allows Ward's method to find groups in multivariate Euclidean space, the same used by Multi-Dimensional Scaling (MDS, also known as Principal Coordinate Analysis). MDS reduces the dimensionality of a distance matrix to represent variances and similarities within data on an ordinance diagram. A set of uncorrelated, orthogonal axes are produced, where each axis created by the MDS procedure has an associated eigenvalue, whose magnitude indicates the amount of variation captured by that axis. MDS is mathematically similar to principal component analysis (PCA). PCA seeks to reduce dimensionality between different variables, also producing sets of orthogonal axes refereed to as components. For both techniques, the number of axes produced equals the number of objects in the distance matrix (which, in our case is the number of \hii regions in the sample; in PCA it would be the number of variables considered). Each axis' relative eigenvalue gives the importance of that axis for summarising the variances in the distance matrix. 

For the input distance matrix $D$, used for the hierarchical clustering, MDS reveals that the first two principal coordinate axes account for 60\,\% and 30\,\% of the total data variance, respectively. A further 7\,\% is represented by axes 3, hence the first three axes account for 97\,\% of the data variation. The subsequent axes each capture less than 1\% variation individually; collectively making up the remaining 3\% variation. The first two principal coordinate axes are shown in Fig.\,\ref{fig:mds} - top, and the second and third coordinates are shown on the bottom. If the original $T_{AD}$ dissimilarity matrix was used as the input to PCA, the resulting PCA ordinance and eigenvalue scores would be equivalent to what we obtain using MDS in Euclidean space. Whilst PCA has a natural parametrisation of eigenvalues, corresponding to some combinations of the input variables, there is not such a correspondence in MDS. The input to MDS is a distance matrix rather than a set of variables, which in our case constitutes distances between the shapes of the \hii regions. Hence, here we use MDS ordinance to see if the results from hierarchical clustering delineate groups that visually correspond to groupings in the MDS Euclidean space. We can then look for associations between observables and the ordinations along the MDS axes. 

Figure.\,\ref{fig:mds} shows that there is a clear correspondence between MDS axis 1 and the groups from hierarchical clustering. Group 2 (red triangle) which represents the outlier region is seen to be ordinated away from the rest of the data in each of the first three axes. Excluding group 2, the ordering of the groups along axis 1 matches the ordering in the CDF plot shown in Fig.\,\ref{fig:clust_cdf}. This suggests that the most variance in the data can be explained by the amount of high curvature points within each group. Groups 6 (pink inverted triangles) and 7 (yellow crossed circles) have the highest values along axis 1. Group 4 (blue crosses) are close to the origin of axis 1, but show a high variance along both axes 2 and 3. Groups 5, 3 \& 1 (cyan diamonds, green plus signs and black circles, respectively) have some cross over along axis 1. There is also cross over between these groups and MDS axis 2, however, with regards to groups 5 (cyan diamonds) and 3 (green plus signs), which are shown to have similar CDFs and originate from the same parent branch on the dendrogram, MDS axis 2 reveals that there is a distinct separation of the groups along this axis. The variance along axis 2 is likely due to another feature of the curvature distributions, since regions with high scores on axis 2 are at the extremes of axis 1. Groups 5 (cyan diamonds) and 4 (blue crosses) have similar scores on axis two, but they appear separated along axis 3 with some cross over around the origin of this axis. There is no further correspondence with group and axis 3, that is not already apparent from axes 1 and 2.

We have shown from the visual inspection of the groups, the group CDF plot and the MDS ordinances that the hierarchical clustering procedure is identifying groups that share similar morphological features. Both the group CDF plot and MDS ordination highlight quantitatively the similarities between groups 3 and 5, yet visual inspection of group 4 reveals a range of shapes are included. We therefore must consider the reliability of the groups which thus far have been identified by an arbitrary cut at a given height on the dendrogram.

\subsubsection{Group \& Method Validity} \label{sec:group_valid}

As outlined at the start of this section, we must ensure that the extraction of shape from our \hii regions is systematic and repeatable. Furthermore, a future use of our unsupervised clustering method is to identify a training set for use in machine learning supervised classification of \hii regions. We have hence carried out a number of robustness checks on both the shape data and performance of the analysis method to determine how well the method copes with different selection choices.

Firstly, in order to test how reliable the identified groups are, we applied multi-scale bootstrap re-sampling \citep{shimodaira2004approximately} to the data. This provides a $p$-value for each merge on the dendrogram, which represents the probability that each respective merge is intrinsic to the data. In standard bootstrap re-sampling, the original data of size $n$ is considered, and bootstrapped data-sets (also of size $n$) are generated by sampling data from the original set with replacement. By applying the same hierarchical clustering procedure to the bootstrapped sets, one can compare how many times a group from the original data appears in a bootstrap replicated dendrogram. However, by always considering a sample size of $n$, biases can be introduced to the bootstrap probabilities \citep{shimodaira2004approximately}. Multi-scale bootstrap re-sampling generates a range of bootstrapped data-sets, which can be smaller or larger than $n$. For each sample size, hierarchical clustering is applied, and the number of times a given group from the original data appears in a replicated dendrogram is counted. This is repeated a number of times at each sample size, resulting in an approximately unbiased $p$-value for each group in the original data. For our distance matrix $D$, the largest group obtained from our sample size of $n = 76$ contained 28\% of the data. We hence set the lower bound of the multi-scaling to $0.5 \times n$ and increased in steps of $0.1 \times n$ to an upper-bound of $1.4 \times n$, at each step using $10^4$ bootstrap iterations. This set up yielded reproducible $p$-values, where all $p$-values above 90\% had a standard error less than 0.6\%. By standard normal theory this equates to a confidence interval on the $p$-values of $\pm$1.2\%. The $p$-values for each merge are shown in Fig.~\ref{fig:dendrogram} in red above each group outlined by the constant cut and precluding merges higher up the diagram. From left to right, the associated $p$-values for the 7 previously defined groups are: Group 7: 51\%, group 6: 78\%, group 2 is the outlier object, which has a $p$=88\% to be joined to group 1, group 1: 94\%, group 4: 94\%, group 5: 94\%, group 3: 97\%. If groups 7 and 6 are joined, the respective $p$-value for all of these objects to be within one group is then 90\%. Hence, for the discussion of group properties and parameters, we consider groups 1, 4, 5, 3 and merge the two regions from group 7 into group 6. 

We then ran some further checks to determine how well the group structure of the dendrogram remains when varying certain selection choices pertaining to the shape data. The first of these was the initial sigma value used to extract the boundaries of the \hii regions. Our reasoning for using the 1$\sigma$ contour level was explained in Sect.\,\ref{sec:data}, with lower $\sigma$ levels including too much noise and higher levels not capturing enough detail. Here we test whether smaller changes to the sigma level affect the results. The analysis was re-run with 0.8 and 1.2$\sigma$ contours and in both cases, some of the group associations obtained from the 1$\sigma$ data changed. This is explained by the fact that changing the threshold level inherently changes the shape considered. The extent to which the shape changes differs on an individual basis, with some regions much more susceptible than others. In turn, the computed pairwise distance matrix representing the region shapes also changes, which can have a knock-on effect for the hierarchical procedure. This is due to how the inter-group distances are computed by the agglomeration method, which considers each object's pairwise distance within a given group. In our case, the Ward method is less amenable to small within-group distance changes than, for example, the single-linkage method, as used in friends-of-friends algorithms \citep{ferreira2009comparison}. Nevertheless, larger differences in the distance matrix from those regions whose shape changes substantially will be captured by the agglomeration process, and the extent to which this affects the results would be apparent in the final groupings.

To determine whether lowering or increasing the $\sigma$ level has more affect on the shape and resulting groupings, we first computed the bootstrapped $p$-values for the dendrograms of the entire $\pm$0.2$\sigma$ data. In each case, they were notably lower than for the 1$\sigma$ data with mean values of 71\% for 1.2$\sigma$ and 84\% for 0.8$\sigma$, compared to 93\% for the 1$\sigma$ data. This suggests that the groups outlined at these levels are less credible than the 1$\sigma$ level. In principal, one could test various sigma levels for resulting groups with the optimum $p$-values, and we will employ this approach in our follow-up work. Next, we switched the shape data of randomly selected individual regions to 1.2 or 0.8$\sigma$ and repeated the analysis. On average, 18\% of regions changed groups at 1.2$\sigma$ and 29\% of regions changed group at 0.8$\sigma$. Although the 1.2$\sigma$ data was less likely to result in a individual region changing group, the lower overall $p$-values for this level may be explained by these shapes being more smoothed with fewer features, so the pairwise distances are generally lower and it is more likely for them to be grouped with different similar regions. For the 0.8$\sigma$ regions, the larger fraction of individual regions changing group is expected since lowering the threshold allows for more detail and extended features with high curvature values to be captured. In all cases where a region changed group, the rest of the group structure from Fig.\,\ref{fig:dendrogram} remained. In addition, when a change was observed, it was to neighbouring branches, and no region moved in to group 6 from the right hand side of the dendrogram. This confirms that the hierarchical groupings are not sensitive to small changes in the distance matrix, and that we may be able to quantify how and when a region moves to/from a given group. This will be investigated further in Paper II.

The way in which the \textquoteleft edge' of the \hii region is defined in order to extract its shape must hence be carefully considered when conducting this type of shape analysis. Although in our case, the 1$\sigma$ level appears to be the best representation, this may not be the case for other surveys with different noise profiles.

A similar test was carried out for the spline knot spacing. The spacing is in essence the resolution of the shape data, hence changing this would also in some cases change the description of the region shape, due to finer details either being included or smoothed out. We initially used a spacing of 0.54\,pc, corresponding to the beam size of the survey at a far distance of 19.2\,kpc. Hence, changes to a region's distance subsequently changes the knot spacing along its boundary. Smaller distances have larger spacings (lower resolution) and vice versa. We changed the distances of regions in the sample by a random amount between $\pm$20\% to address both issues of the sample distance errors and how different resolutions affects the results. We found that when decreasing the distance/resolution of a region, 18\% changed groups and when increasing the distance/resolution, 53\% of regions changed groups. Capturing more detail of the shape thus has a much more profound affect than smoothing the data. This agrees with the results from the MDS ordination that shows a correlation between assigned group and number of high-curvature points along the region boundary. The fact that increasing the spline resolution has the most notable affect on the groups supports our choice of using the far distance resolution to reduce distance biases, which is further discussed in Sect.\,\ref{sec:group_dist}. Whilst this is an important limitation to identify, we do note that changing the resolution of individual regions in this manner affects the systematic nature of the shape analysis. It is hence expected that changing the shape descriptor of a region by a substantial amount will cause it to move to a different group associated to another morphology class. Furthermore, certain regions of a given group each moved to the same group after changing the resolution. For example, each region that moved from group 3 was relocated to group 1 after increasing the resolution. This suggests that we may be able to further quantify the morphological groupings with a more controlled sample. 

In terms of our sample of \hii regions, the region images in Appendix\,\ref{app:group_images} show that we did arrive at a varied selection in terms of morphological features. To test whether our group results are sensitive to different samples, we removed a number of regions from the data and re-run the analysis. 18, 25, 32 and 38\% of regions were removed at random and in each case, the resulting group associations matched that of Fig.\,\ref{fig:dendrogram} for the remaining regions. Similarly, we tested whether the group structure remains when removing all regions belonging to a given group. Again, all of the group structure is kept intact in the absence of any given group. These results suggest that the six groups we identify do in fact represent distinct categories of morphologies, and is promising for the prospects of a producing a training set from synthetic data in future work. 

\section{Discussion}
\label{sec:discussion}
In this section, we consider the physical properties of the \hii regions within each group. We look for any associations between these parameters and assigned group and investigate whether it is some physical property responsible for the 30\% variation along axis 2 of the MDS ordination, having already shown the 60\% variation represented by axis 1 is likely linked to the amount of high-curvature values within each region. We also discuss in this section the outliers identified by hierarchical clustering and investigate assumptions made about the assigned distances and ambient densities. We also outline potential future applications of our statistical methodology to larger samples of \hii regions, and other diffuse astronomical objects, and how we intend to further quantify the statistical clustering using synthetic observations. 

\subsection{Region properties by group} 
\subsubsection{Position, Radius and Distance} \label{sec:group_dist}

\begin{figure}
	\includegraphics[width=\columnwidth]{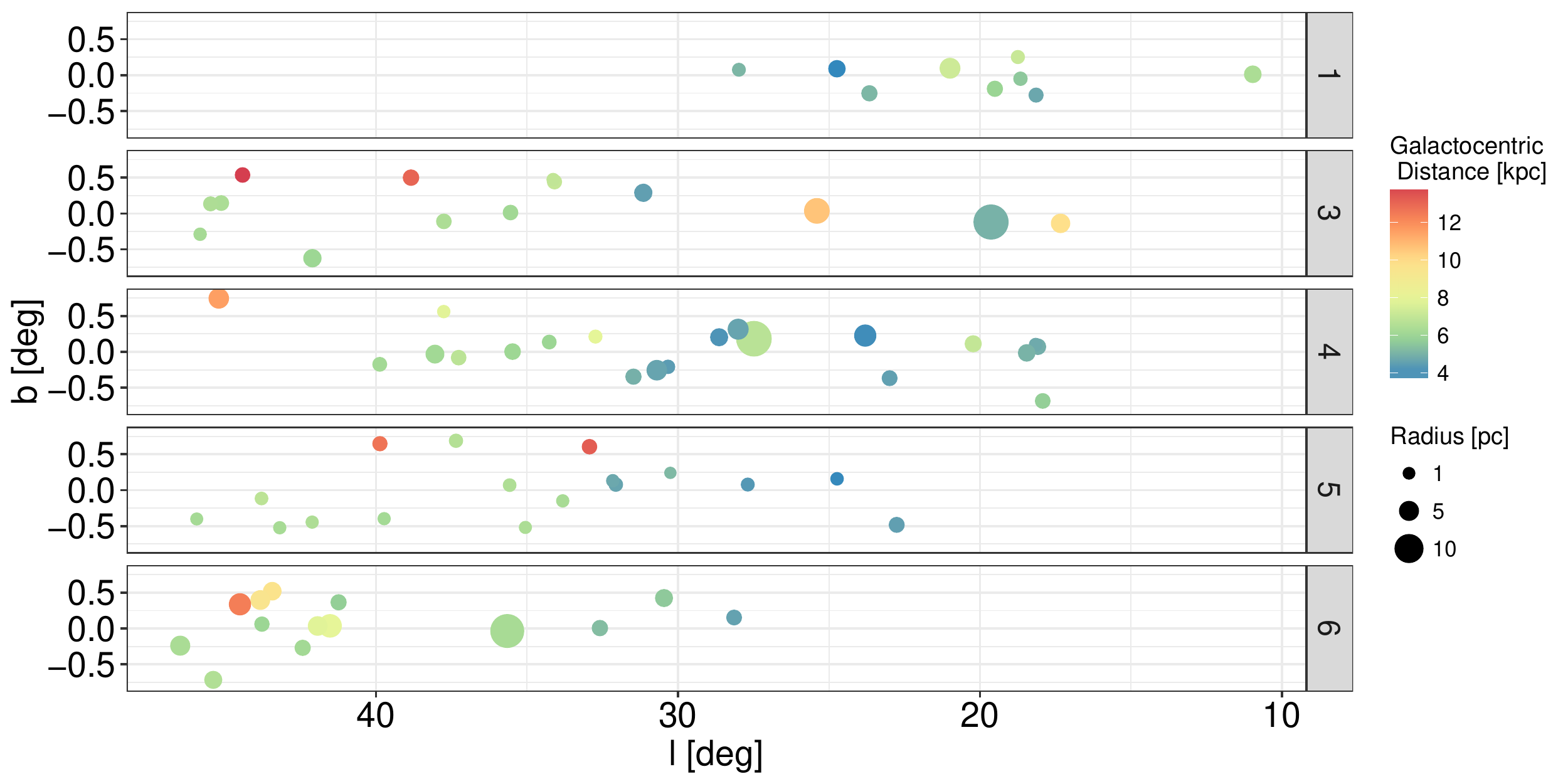}
	\caption{Galactic position of our sample of \hii regions, split by group. Points are coloured by their Galactocentric distance and scaled linearly with their physical effective radius. Note that the two regions in group 7 from Fig.\,\ref{fig:dendrogram} have been merged into group 6 and the outlier region in group 2 is not shown.}
	\label{fig:pos_clust}
\end{figure}

Figure~\ref{fig:pos_clust} shows the positions of the \hii regions -- Galactic longitude and latitude, with points coloured by Galactocentric distance -- split by assigned group. Group 1 only contains regions with $l<30$\dg and group 6 only has regions with $l>28$\dg, however this is not exclusive, with groups 3, 4 \& 5 containing regions with a range of $l$ values. There are no group preferences for regions with positive or negative latitude values, nor large or small Galactocentric distances. This suggests that the various shapes of \hii regions located in the first half of this quadrant of the Galaxy are homogeneously distributed in Galactic latitude, with a small preference for those at a given Galactic longitude to share a common morphology.

\begin{figure*}
	\includegraphics[width=\textwidth]{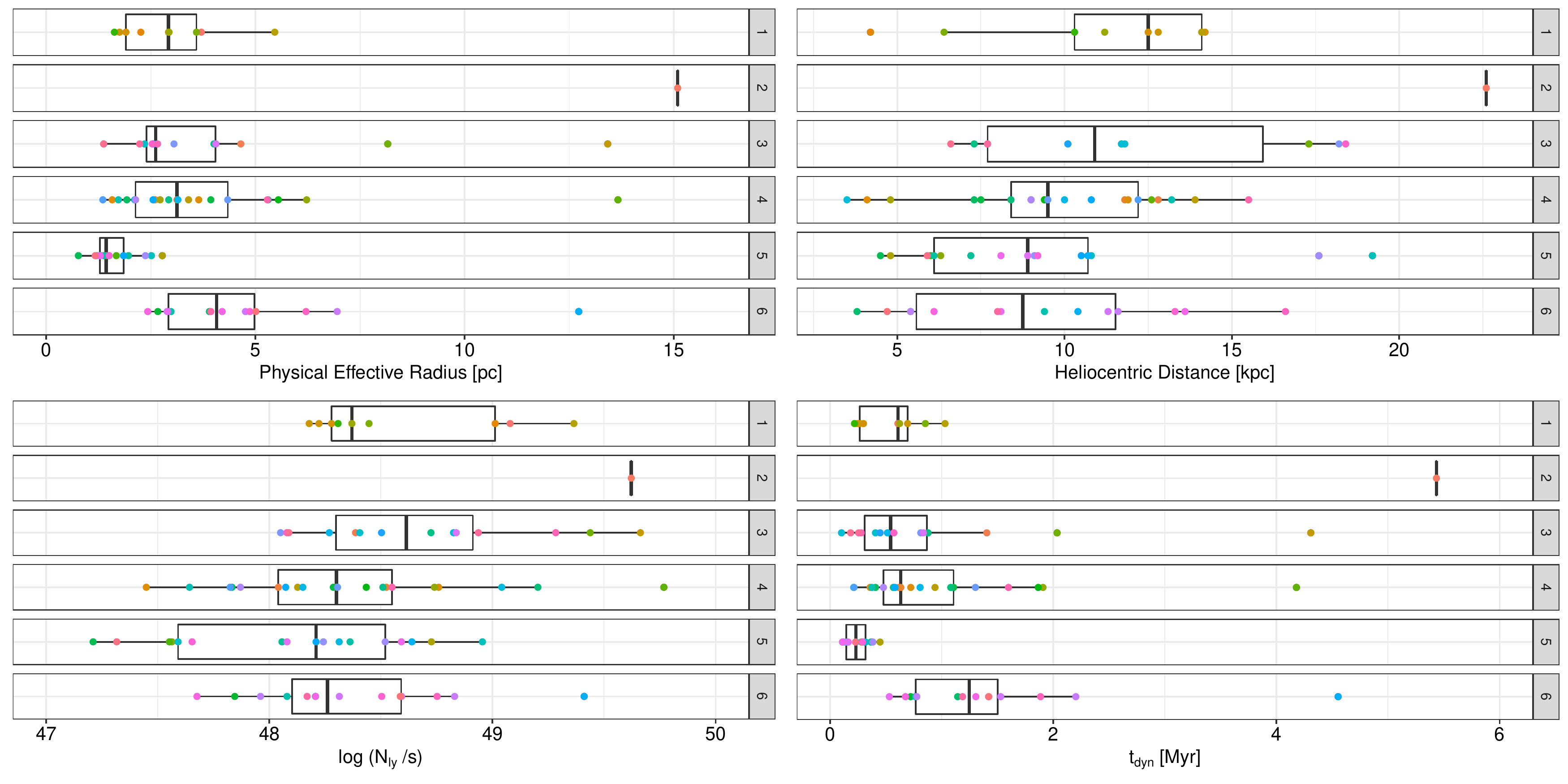}
	\caption{Summary of \hii region properties by group. Top left shows physical effective radius, top right shows heliocentric distance, bottom left shows the number of ionising photons within the boundary of each region and bottom right shows the dynamical age. Points are consistently coloured by the region's Galactic longitude, allowing for identification of corresponding regions across plots. As with Fig.\,\ref{fig:pos_clust}, the two regions from group 7 are merged into group 6. }
	\label{fig:bp_summary}
\end{figure*}

Figure~\ref{fig:bp_summary}, top-right, shows the distribution of \hii region heliocentric distance by group. As previously outlined, we sought to ensure that no bias was given to regions at a near distance. This intent is confirmed by Fig.~\ref{fig:bp_summary}, with no preference for regions at any given distance appearing in a particular group. Further to our tests using different spline intervals (Sec.\,\ref{sec:group_valid}), we tested whether considerably smaller intervals (equal to 0.10\,pc and 0.29\,pc), that captured much more curvature detail for the closer regions, would show a bias in the clustering results and found that, in these cases, there was a preference for objects at greater distances to be sorted into the same groups. This preference was most profound at the smaller interval of 0.10\,pc, where 72\% of regions at a distance greater than 12.5\,kpc were placed in one group. For the interval of 0.29\,pc, 39\% of the far regions were then placed into one group. Since there is no preference for the far regions, nor those at the nearest distances, to be placed in a given group at the 0.54\,pc spline interval used, we can infer that any associations between region parameters and groups is not associated to any distance/angular resolution bias. 

We do find associations between the region radii and assigned group (Fig.\,\ref{fig:bp_summary}, top-left). Since the curvature measurements of each region concern the size-and-shape, this is not an unexpected result. However, effective radius as a measure was not a direct input to the hierarchical clustering, since the curvature distributions or average curvature values per region do not always correlate with the effective radius. Therefore, the fact that our results have the following associations means that this size information is in some cases retained by the curvature distributions: 59\% of regions with a radius less than 1.6\,pc are assigned to group 5, and 53\% of regions with a radius greater than 4\,pc are assigned to group 4. Because of this association between large regions and group 4, we cannot attribute size as the primary reason for the variance along axis 2 of the MDS ordinance, since group 4 has low scores along this axis. The \hii regions physical size can be linked to both ionising mass and age, for an assumed constant medium density. We therefore estimated the amount of ionising flux from each region, to provide estimates for the stellar masses, which when compared to the size of the regions, can also give estimates for the region dynamical ages.

\subsubsection{Lyman Continuum Flux}
Using the integrated radio continuum flux density from within the boundary of each \hii region, we estimated the number of Lyman continuum photons from the following equation \citep{1976AJ.....81..172M}, derived from the model developed by \citet{1967ApJ...147..471M} and assumptions outlined by \citet{1968ApJ...154..391R}:

\begin{equation}
	N_{ly}=7.54\times 10^{46}\left(\frac{S_\nu}{\rm Jy}\right) \left(\frac{D}{\rm kpc}\right)^2 \left(\frac{T_e}{10^4\rm K}\right)^{-0.45} \left(\frac{\nu}{\rm GHz}\right)^{0.1}
	\label{eq:nly}
\end{equation}

\noindent Where $S_\nu$ is the measured total flux density, $D$ is the distance to the Sun, $T_e$ is the electron temperature and $\nu$ the frequency of the radio emission. We assume all regions are optically thin and have an electron temperature of 8,500\,K. For $S_\nu$, the intensity values from the MAGPIS images are given as Jy/beam, we therefore extracted the total emission by first determining the beam integral (1.133 $\times\,beam \,size^2 \simeq 38.1''$) and dividing by the pixel scale, which gives the number of pixels in the beam ($\simeq 9.5$\,pixels). Then, dividing the sum of the pixel values within the boundary of each region by the number of pixels in the beam yields the flux density in Jy. 

Derived $S_{\nu}$ and estimated $N_{ly}$ values for each of the 76 \hii regions are listed in Table\,\ref{tab:summary_table} in Appendix \ref{app:table}. These ionising flux estimates are lower limits due the assumption of ionisation bound regions. A small amount of ionising flux can escape into the outer photo-dissociation region, and some flux may be attenuated by dust within the region. After identification of the probable ionising sources for a sample of \hii regions, \citet{2008ApJ...681.1341W,2009ApJ...694..546W} estimate that the $N_{ly}$ values calculated from the MAGPIS images are a factor of 2 lower than the expected flux from the ionising stars. We do not correct for this in our calculations and instead assume all inferred masses are lower-limits. Since this is systematic across the sample, estimated values are still representative of the sample distributions. 

\begin{figure}
	\includegraphics[width=\columnwidth]{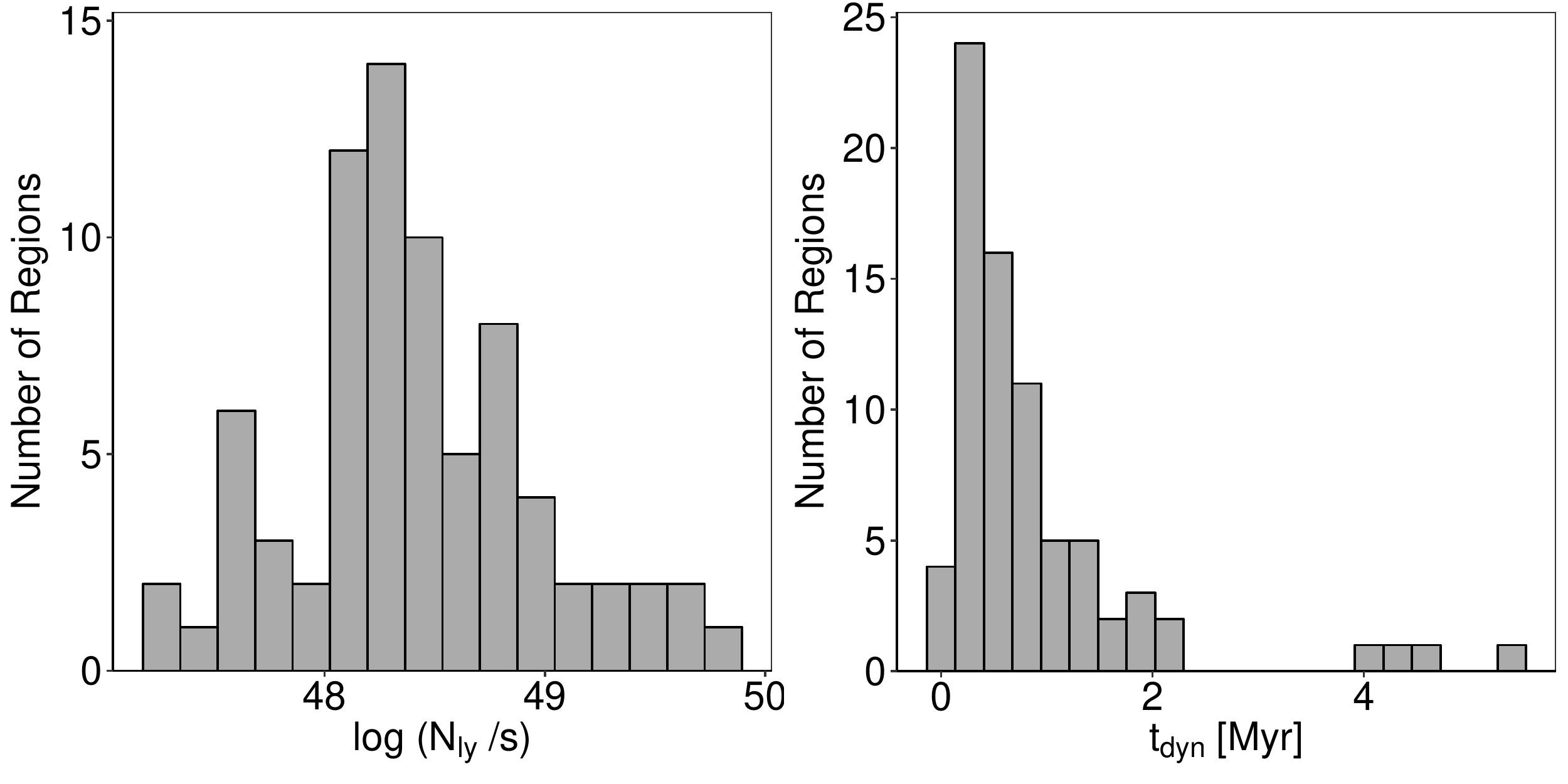}
	\caption{Distribution of the number of ionising photons within the boundary of each \hii region (left) and the dynamical age of each region (right).}
	\label{fig:nly_thii_hist}
\end{figure}

The distribution of $N_{ly}$ is shown in Fig.~\ref{fig:nly_thii_hist}, left. The log $N_{ly}$ values range from 46.92 to 49.77, corresponding to single ionising stars of spectral type B0.5V and O5V, respectively (Table $\rm II$ in \citet{1973AJ.....78..929P}), with a peak of $\sim$\,48.3 corresponding to an O9V star. This distribution further confirms the expectation from C06 that many of the bubbles are the result of ionisation and stellar winds from late O and early B type stars. This was also found by \citet{2010ApJ...709..791B} for a selection of the C06 bubbles, however, for the bubbles that are common between our sample and theirs, we note a higher estimate of $N_{ly}$, due to not assuming the near distance from the kinematic distance estimates. 

\citet{2010MNRAS.401..275W} found that for star cluster masses exceeding $10^2$\,\Msolar, random sampling from the stellar initial mass function is highly unlikely, and that the star cluster mass is related to the mass of the most massive star it hosts. For a star cluster of $10^2$\,\Msolar, the corresponding most massive star has a mass of $\sim$ 8\,\Msolar. Since a single O9V star has a mass of $\sim$ 19\,\Msolar \citep{2010A&A...524A..98W}, and assuming that the majority of the ionising flux within each \hii region is the result of the most massive star within the cluster, we can hence determine the minimum star cluster mass required for each \hii region. 

Figure\,\ref{fig:bp_summary}, bottom-left, shows the distribution of $N_{ly}$ values by group. There is no apparent association between group and the region's number of Lyman photons, with the most massive regions distributed across all groups except group 5. The largest value of log $N_{ly}$ in group 5 is 48.9, which corresponds to a single O6.5V star. From \citet{2010A&A...524A..98W}, an O6.5V star has a spectroscopic mass of $\sim$ 30\Msolar, which would require a cluster of $\sim$ $10^3$\Msolar to produce \citep{2010MNRAS.401..275W}, this is hence the upper mass limit for group 5. Group 1 contains only intermediate- to high-mass regions, with a lowest log $N_{ly}$ value of 48.2, which is between an O9V and an O9.5V star, corresponding to a minimum cluster mass of $\sim 10^{2.5}$\Msolar for this group. In both groups 1 \& 5, there appears to be a break in the parameter space between intermediate and high mass clusters, and lower and intermediate mass clusters, respectively. However, splitting these groups each into two subgroups at their highest merge levels on the dendrogram in Fig.\,\ref{fig:dendrogram} does not delineate the groups by their respective $N_{ly}$. Although the group $p$ values for these subgroups are all > 92\%, it would not be the cluster mass in this case that warrants separating out the two groups. Instead, we take these two groups as representing their respective $N_{ly}$ distributions, with no high mass regions in group 5 and no low mass regions in group 1. The remaining groups 3, 4 \& 6 all show a large spread in $N_{ly}$ values, with groups 4 and 6 each containing one region outside of their respective group's box-plot tails. 

\begin{figure}
	\includegraphics[width=\columnwidth]{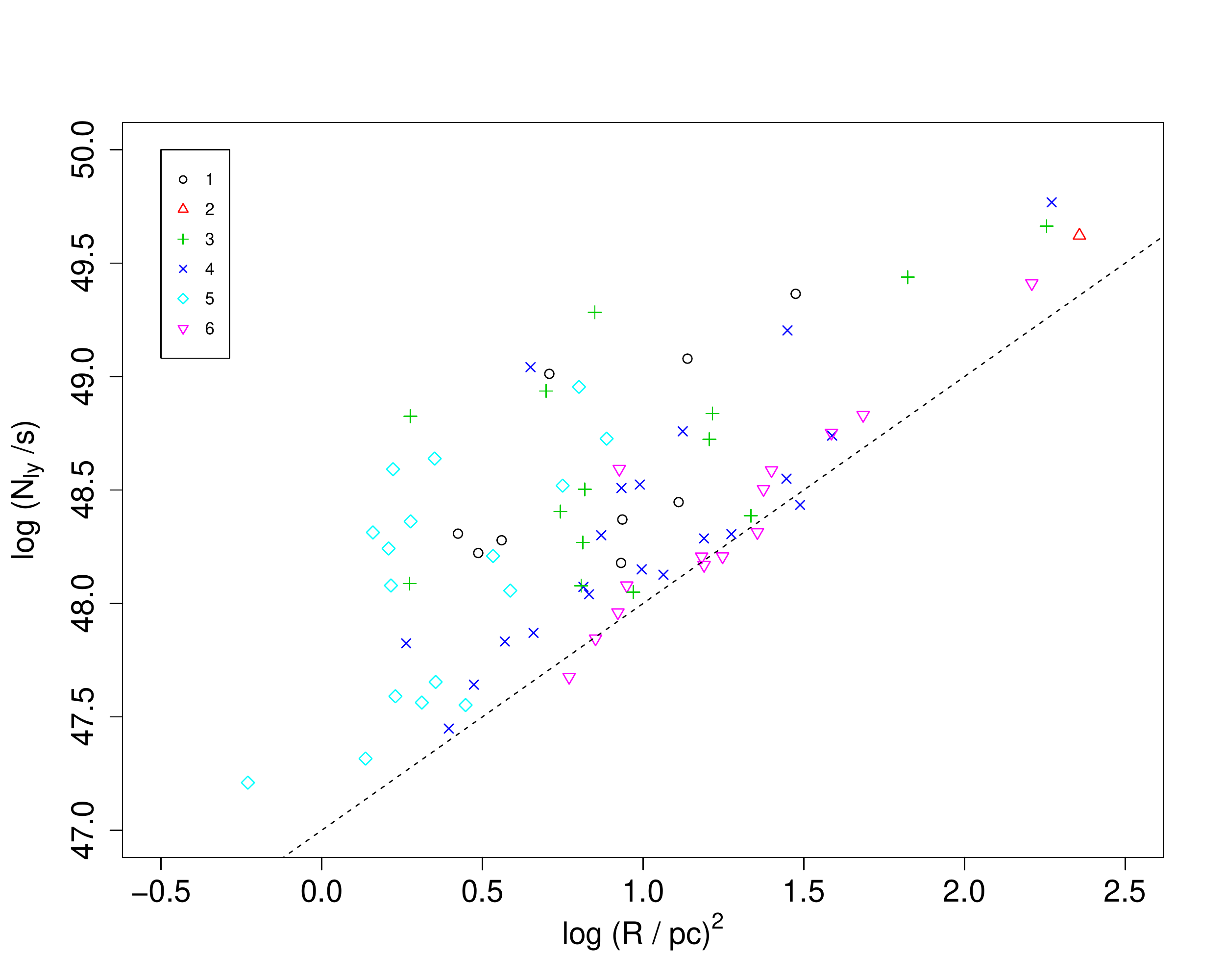}
	\caption{Physical effective radii of each \hii region versus the number of ionising photons. Colours and symbols correspond to the groups identified from hierarchical analysis. A constant surface brightness of 1$\cdot$10$^{47}$ photons\,/\,s per pc$^2$ is indicated by the dashed line.}
	\label{fig:surf_bright}
\end{figure}

Figure\,\ref{fig:surf_bright} shows each region's $N_{ly}$ values versus physical surface area. There is a clear power-law cut off in the parameter space, indicating a limiting surface brightness of the MAGPIS data. For reference, the dashed line in Fig.\,\ref{fig:surf_bright} corresponds to a surface brightness of 1$\cdot$10$^{47}$\,photons/s per pc$^2$. Colours/symbols of points match those from the MDS plots in Fig.\,\ref{fig:mds} and represent the identified groups. All but one of the regions within group 6 appear to be at the detection limit of the survey, suggesting that for this group, the noise may be interfering with what we identify as the \textquoteleft edge' of the \hii regions. This is a possible explanation as to why group 6 appears isolated in Fig\,\ref{fig:dendrogram}, not merging with the rest of the data until a the largest height on the dendrogram. The fact that these objects have been distinctly grouped by the shape analysis method suggests that regions in the other groups that are also close to the limiting surface brightness are still at a sufficient signal/noise ratio that the edge has been correctly identified. Hence, any inferences made regarding the regions in group 6 may not be as credible as those from the rest of the groups. This is because there may be a larger error in the determined diameter of the regions, and the regions may in fact belong in different groups if their edges were better defined.

\subsubsection{Dynamical Age}
Having determined $N_{ly}$ for each \hii region, their dynamical ages can be determined from the following equation \citep{1968ITPA...28.....S,dyson_williams1980}:
\begin{equation}
t_{dyn}=\left(\frac{4\, R_s}{7\, c_s}\right) \left[\left(\frac{R_{\rm\hii}}{R_s}\right)^{7/4}-1\right]
\label{eq:thii}
\end{equation}

\noindent where $R_s$ is the radius of the Str\"{o}mgren sphere ($=\,3\,N_{ly}/4\pi n_0^2\alpha_B)^{1/3}$, with $n_0$ the ambient particle number density, taken as 10$^3$\,cm$^{-3}$, and $\alpha_B = 2.6 \times 10^{-13}$\,cm$^3$\,s$^{-1}$ is the hydrogen recombination coefficient to all levels above the ground level, $c_s$ is the isothermal sound speed in the ionised gas (=\,11\,km\,s$^{-1}$; \cite{2009A&A...497..649B}) and $R_{\rm\hii}$ is the observed radius of the \hii region. Equation \ref{eq:thii} is the result of analytical models that assume pressure equilibrium between the ionized and neutral shocked gas. Whilst this assumption is reasonable for standard \hii regions, external pressure has a larger influence for (ultra- and hyper-) compact \hii regions.  \citet{2012RMxAA..48..149R,2012MNRAS.419L..39R} addresses this by considering the inertia of the shocked gas that is pushed out by the \hii ionization front and present an updated model. The analytical solution of their model is approximately equal to Eq.\,\ref{eq:thii} for a \textquoteleft cold' surrounding medium. We hence make this further assumption for the ambient environment when using Eq.\,\ref{eq:thii}. More recently, \citet{2015RMxAA..51...27R} explored the combined affects of gas pressure, radiation pressure and self-gravity, we refer those looking for a more complete solution to dynamical ages of \hii regions to this work.

The distribution of dynamical ages for our sample of \hii regions is shown in Fig.~\ref{fig:nly_thii_hist}, right panel. There is a peak at $\sim$\, 0.4\,Myr and only nine regions have an age greater than 1.6\,Myr. This age distribution is on average lower than the one recently obtained by \citet{2017A&A...605A..35P}, whose ages were determined by comparing calculated ionised gas pressures to isochrones of 1D simulations by \citet{2014A&A...568A...4T}, by $\sim$ a factor of two. The likely reason for this is the assumptions we make about the constant, cold ambient density (and to a lesser extent, the constant electron temperature assumed when calculating the ionising flux), which may vary with environment across the regions. 

Figure\,\ref{fig:bp_summary}, bottom right, shows the distribution of dynamical age by group. 84\% (16) of the regions older than 1.1\,Myr are assigned to groups 4 and 6, with group 6 only containing regions with ages > 0.5\,Myr. Group 5 shows the smallest spread in ages, only hosting regions < 0.5\,Myr old (22\% (17) of all regions, 51\% of regions < 0.5\,Myr old). Groups 3, 4 \& 6 host regions with ages exceeding the tails of their respective groups. These regions are discussed in more detail in the next subsection. The distribution of ages by group is similar to that of the effective radius. This is because the dynamical age has a stronger dependence on radii than ionising flux, hence regions with a large radius remain as outliers in the age distribution. 

Despite the associations outlined here, the distributions of radius, ionising photons and dynamical age do not match the MDS scores of the regions along axis 2. This means that we cannot attribute the 30\% variation in the distance matrix that axis 2 represents by any one physical parameter. It is likely due to another feature of the curvature distributions that is not clear from the CDF plots. Since the A-D test is sensitive to the tails of the distributions it may be some subtle differences inherent there. Nevertheless, the absence of low- and high-mass regions in two groups, the result that one group only has small, young regions and that one group is at the surface brightness limit of the survey has all been revealed by the shape analysis method presented here. This shows good evidence that there is a link between shape and physical properties of the regions/environments. It is also promising for using these results as a training set for future data sets.

\subsection{Outliers}\label{sec:outliers}
The apparent outlier from the hierarchical clustering is the single region in group 2. \hii region G012.429-00.049 is the most distant region in the sample (d\,=\,22.6\,kpc), and from its line of sight radial velocity, is outside the Solar circle and hence does not have a near kinematic distance (V$_{lsr}$\,=\,-18.4\,km~s$^{-1}$). This region has the sixth largest angular diameter in the sample, which - combined with its far distance - leads to it having the largest spatial diameter in the sample ($\sim$ 30\,pc). This meant that this \hii region had by far the largest number of interpolation points along its boundary. This may be the reason for this \hii region having a large A-D test statistic score when compared to all other regions. However, given its size and distance, this \hii region is the oldest in the sample at 5.4\,Myr. It may simply be because of how evolved the region is that it has been selected as an outlier from the rest of the sample. We see from the region image in Appendix\,\ref{app:group_images} that the MIR bubble is not in a complex, and the radio continuum emission at the 1$\sigma$ boundary extends further than the 8\mic shell. This is not common to the sample and suggests a potential projection contamination with another radio source. In the statistical study of Spitzer bubbles by \citet{2014MNRAS.438..426H}, they match this MWP bubble to the same \hii region we do, from \citet{1987A&A...181..378C} and \citet{1989ApJS...71..469L}. There is also a separate UC\hii region within this bubble, listed in \citet{1989ApJS...69..831W}. 

Further outliers have been identified from Fig.\,\ref{fig:bp_summary}, assuming that the region shapes are linked to the physical parameters discussed. Region G027.476+00.179 in group 4 has a determined age of 4.2\,Myr and also the largest value of $N_{ly}$. This far exceeds the interquartile range (IQR) of the respective properties of regions in group 4. The distance of this \hii region is quoted as 12.6\,kpc, which is the far kinematic distance and is a very secure assignment \citep{2012ApJ...753...62J}. Similarly, an outlier in group 3, region G019.629-00.095, has an age of 4.3\,Myr, which was also determined with the far kinematic distance of 11.7\,kpc and is another very secure distance assignment \citep{2004ApJS..154..553S}. The other outlier in group 3, region G025.397+00.033, has an age of 2.0\,Myr at a distance of 17.3\,kpc which is outside of the Solar circle. Since in each of these cases, there is little evidence for or no KDA, we suggest that these could be examples where the incorrect assumptions have been made regarding the ambient density or temperature. For example, the images of region G025.397+00.033 in Appendix\,\ref{app:group_images} show an elongated area of radio emission, extending away from the centre of the MIR bubble. If a lower density of n\,=\,500\,cm$^{-3}$ is assumed for this region, its age is then calculated as 1.4\,Myr, which is just within the upper extreme of that group's age values. Group 6 hosts the second oldest region with the fifth highest $N_{ly}$, G035.649-00.053 (N68 from C06), which at a far kinematic distance of 10.4\,kpc is calculated as 4.6\,Myr old. As discussed previously, group 6 has been identified as regions on the limit of surface brightness detection, hence there is a larger error in the diameter determination of these regions which would influence the calculated dynamical age. 

\subsection{Future Applications}

When testing the shape analysis functionality of our method, regular shapes such as circles, ellipses and regularly perturbed closed curves were compared to the observed sample. In each case, the test regions were ordinated far away from the observed regions on the MDS plots, and were grouped in to their own branches on the dendrograms. Hence, an application of this analysis method would be to test the efficacy of synthetic observations of numerical simulations of \hii regions. This will be the main focus for Paper II, with preliminary results showing that modern simulations do indeed produce synthetic images that are comparable to the observed data. Hence, for varying initial conditions and at different simulated ages, we can determine when the synthetic regions appear in the groups that we identify. Furthermore, we can take advantage of these controlled parameters to better quantify our groups and find out when a region would move between groups. We have already seen evidence for regions moving systematically between groups from our resolution testing. Ultimately this would help lead to a supervised classification scheme of \hii regions based on their shapes, whereby each class is associated to a specified range of physical properties.

A further use of the synthetic observations is to consider different projections of a region at a given stage, to see how this affects the obtained shape of the region. Although in the present work, we consider a variety of \hii region shapes, it remains to be seen from this alone whether or not the line of sight angle to an individual source would have a large influence on its shape. Through the testing of our analysis, we saw that a region was more likely to move between groups if its shape changed in a definite manner, such as more detail being captured at a lower contour level, or a higher spline resolution. We can hence speculate that if no greater amount of fine detail was revealed by a change in projection, then the region should remain in the same group when viewed at different angles. This will be discussed in detail in Paper II. 

There are various other diffuse astronomical objects that share common morphologies, such as planetary nebulae, supernova remnants, giant molecular clouds, and cold molecular clumps. If the shape data of these objects can be extracted in a systematic manner like we have done for the \hii regions in this study, our analysis method could be applied to these objects. Due to the curvature shape descriptor and the non-parametric statistical tests used in our method, the regions do not necessarily have to be closed contours. 

\section{Conclusions}
\label{sec:conclusions}

We performed an unbiased shape analysis of a sample of 76 \hii regions by analysing the curvature distributions of the regions boundaries. These were obtained systematically at a constant signal level above the background noise in each image from the MAGPIS radio continuum data. By applying hierarchical clustering and multi-scale bootstrap re-sampling to the data, we obtained six groups, delineating \hii regions of similar morphologies. This was confirmed by visual inspection of the images \footnote{As shown in Appendix \ref{app:group_images}, available in the online version of this paper.} and quantitatively by the ordinance technique of multidimensional scaling. We found that 97\% of the variation in the shape data is represented by three principal coordinate axes, 60\% of which is likely due to the amount of high curvature points along a region boundary, i.e. the level of curvature perturbation. Investigation into the association of physical parameters and the group assigned by our methodology revealed the following results: 

There was little association between the region position in the Galaxy and assigned group, with objects at varying Galactocentric distance and Galactic latitude appearing in all groups. We found that one group contained regions with Galactic longitude < 30\dg and another had only regions with $l$ > 28\dg. However, the remaining three groups contained regions with the full distribution of $l$ values. This suggests that for this section of the Galactic Plane, \hii region shape is homogeneously distributed across Galactic latitude, with a small preference for those at similar Galactic longitudes to share a common morphology.

One of the six groups contained only small (<1.6\,pc), young (<0.5\,Myr) \hii regions. This was not exclusive, with 59\% and 51\% respective completeness. It does show that the size-and-shape information contained within the curvature distributions and the statistical clustering of the data reflects this common feature for this group. There was also a preference for this group to contain only low- to intermediate-mass ionising clusters (maximum of $\sim$ 10$^3$\Msolar). There was a further preference for ionising mass in another group that contained only intermediate- to high-mass clusters (minimum mass $\sim$ 10$^{2.5}$\Msolar). There was no further preference for regions of a given ionising mass to be placed in any particular group. Using these results, we discussed five outliers in the sample and looked at whether there were possible projection effects or incorrect assumptions made as to the ambient density or electron temperatures. In each case there was either little evidence for or no kinematic distance ambiguity. 

One of the identified groups was distinctly separated from the rest of the data by our analysis. It was revealed to contain regions that were at the surface brightness detection limit for the survey. This suggested that the noise may be interfering with our extraction of the boundary of these regions, and that the shape of these objects is less accurately represented than in the rest of the data. For a deeper survey, these regions may in fact belong to a different group in the analysis. The groupings we obtain remain present if this (or any individual) group is removed from the data set. The group associations also remain if a random number of regions are removed from the data. This further illustrates that common morphological features were readily identified by our method. 

It was evident from testing the methodology that certain selection choices affect the resulting group structure from the hierarchical process. We found that lowering the threshold level used to define the \textquoteleft edge' of the \hii region resulted in higher curvature portions of the shape to be included in its descriptor. In some cases this was due to the noise in the images, in others it can be attributed to structure of comparable scale to the resolution used. This had a much larger influence on the group results than increasing the threshold value, concurring with  the MDS result of the number of high-curvature values along the boundary having a large influence on the groupings. Increasing the threshold produced generally more smoothed edges with less defined detail. The $p$-values (from the bootstrapping) associated with the groups obtained in each case were lower than those obtained for the 1$\sigma$ value used. For our future work, we will use this approach to identify the optimum threshold level to use, for varying noise profiles that can be readily controlled with synthetic observations. We will also use the synthetic data to test whether line of sight projection has a significant influence on the observed planar shape.  
	
The largest source of error for the shape analysis was due to the sampling resolution used to obtain the curvature distributions, which is determined for each region from its distance. We discuss why using the spatial sampling resolution corresponding to the image resolution at the far distance (0.54\,pc at 19.2\,kpc) removed any bias that is attributed to nearby regions. Our results show no association between heliocentric distance and group at this resolution. However, as with the threshold levels, we found that changing the regions distance by $\pm$20\% (which in turn changes the sampling resolution) did affect the groups from the hierarchical process. Due to the errors associated with kinematic distance determination, this is a limiting factor of the analysis process. 18\% of group assignments changed when the distances are reduced, which corresponds to reducing the sampling resolution, and 53\% of assignments changed when the resolution is increased. We do note that changes to the shape descriptor of individual regions would inevitably lead to the region being grouped with regions of different morphologies. Furthermore, in some cases, multiple regions from a given group were each relocated to the same new group after changing the resolutions, suggesting that the shape descriptors are affected in a common manner by the resolution change. In our forthcoming work, we will investigate whether these inherent deviations in the description of the shape of \hii regions, which arise from both selection choices and observational errors, can be quantified in a fully systematic way that would allow for a classification scheme to be constructed from shape analysis.

From many levels of abstraction, our results show good evidence for associating \hii region shape to the region's physical parameters, and that shape can be used as an intrinsic measure. With such a large amount of high resolution images readily available, there are many potential applications of this approach to larger samples of \hii regions, and other astronomical objects. We would like to reiterate that for diffuse objects, one must have a clear definition for the shape they extract and execute this in a systematic and repeatable manner. In our following paper, we will compare synthetic observations of \hii regions with given initial conditions and projections to our observed sample, to see if the theoretical values match those calculated from observations and if so, probe how the initial conditions of formation affect the shapes of these regions. This will also allow us to further quantify our groups, moving towards a supervised classification scheme for \hii regions. 

\section*{Acknowledgements}

We thank the anonymous referee for their useful comments and suggestions that have improved this work. JCW acknowledges the studentship provided by the University of Kent. We thank Pedro Palmeirim for the informative discussion regarding the dynamical age distribution of our sample. We also thank Ahmad Ali and Tim Harries for providing us results from their numerical simulations, which allowed us to carry our preliminary work ahead of our follow-up paper.

%%%%%%%%%%%%%%%%%%%%%%%%%%%%%%%%%%%%%%%%%%%%%%%%%%



\bibliographystyle{mnras}
\bibliography{paper_bib} % if your bibtex file is called example.bib

\clearpage
\newpage

%%%%%%%%%%%%%%%%%%%%%%%%%%%%%%%%%%%%%%%%%%%%%%%%%%

%%%%%%%%%%%%%%%%% APPENDICES %%%%%%%%%%%%%%%%%%%%%

\appendix
\onecolumn

\section{Summary Table}\label{app:table}

%\begin{landscape}
\renewcommand{\tabcolsep}{3pt}
\setlength\LTcapwidth{\textwidth}

\begin{longtable}{|c|c|c|c|r|r|c|c|c|c|c|c|c|c|c|}

\caption{\label{tab:summary_table}This table lists the individual properties of all \hii regions investigated in this paper. We list the WISE \citep{2014ApJS..212....1A} and MWP \citep{2012MNRAS.424.2442S} catalogue name of the source, the C06 \citep{2006ApJ...649..759C} name where applicable, the Galactic longitude and latitude, line of sight radial velocity and distances (taken from the WISE catalogue), angular and spatial effective radius, the ionised flux at 1.4\,GHz we determine from within the 1$\sigma$ boundaries, the number of ionising photons, Str\"{o}mgren radius, dynamical age and assigned group number from this analysis. Note that the two regions with group 6* are those originally labelled as group 7 in Fig.\,\ref{fig:dendrogram} and later merged into group 6} \\

\hline
WISE ID & MWP ID & C06 & $l$ & \multicolumn{1}{c|}{$b$} &   \multicolumn{1}{c|}{V$_{\rm lsr}$} & D & R$_{\rm GC}$ & \multicolumn{2}{c|}{R$_{\rm eff}$} & Flux & N$_{ly}$ & R$_{\rm s}$ & t$_{dyn}$ & G \\ 
 &  & & \multicolumn{2}{c|}{[deg]}   &  [km/s] & \multicolumn{2}{c|}{[kpc]}  & [\arcmin] & [pc] & [Jy] & [s$^{-1}$] & [pc] & [Myr] & \\ \hline
\endfirsthead

\multicolumn{15}{c}%
{{\bfseries \tablename\ \thetable{} -- continued from previous page}} \\
\hline
WISE ID & MWP ID & C06 & $l$ & \multicolumn{1}{c|}{$b$} &  \multicolumn{1}{c|}{V$_{\rm lsr}$} & D & R$_{\rm GC}$ & \multicolumn{2}{c|}{R$_{\rm eff}$} & Flux & N$_{ly}$ & R$_{\rm s}$ & t$_{dyn}$ & G \\ 
&  & & \multicolumn{2}{c|}{[deg]}   &  [km/s] & \multicolumn{2}{c|}{[kpc]}  & [\arcmin] & [pc] & [Jy] & [s$^{-1}$] & [pc] & [Myr] & \\ \hline
\endhead

\hline \multicolumn{15}{|r|}{{Continued on next page}} \\ \hline
\endfoot

\hline
\endlastfoot

G010.964$+$00.006 & 1G010965$+$000116 &  & 10.965 & 0.012 & 17.7 & 14.1 & 6.2 & 0.90 & 3.7 & 0.723 & 49.08 & 0.7 & 0.6 & 1 \\ 
  G012.429$-$00.049 & 1G012432$-$000410 &  & 12.432 & $-$0.041 & $-$18.4 & 22.6 & 14.6 & 2.30 & 15.1 & 0.982 & 49.62 & 1.1 & 5.4 & 2 \\ 
  G017.336$-$00.146 & 1G017330$-$001379 &  & 17.330 & $-$0.138 & $-$6.2 & 17.3 & 9.7 & 0.92 & 4.7 & 0.097 & 48.39 & 0.4 & 1.4 & 3 \\ 
  G017.928$-$00.677 & 1G017921$-$006843 & N20 & 17.921 & $-$0.684 & 39.1 & 12.8 & 5.5 & 0.70 & 2.6 & 0.080 & 48.04 & 0.3 & 0.6 & 4 \\ 
  G018.076$+$00.068 & 1G018081$+$000708 &  & 18.081 & 0.071 & 58.2 & 11.8 & 4.7 & 0.91 & 3.1 & 0.288 & 48.52 & 0.5 & 0.6 & 4 \\ 
  G018.144$-$00.281 & 1G018141$-$002783 &  & 18.141 & $-$0.278 & 53.9 & 4.2 & 4.5 & 1.85 & 2.3 & 6.984 & 49.01 & 0.7 & 0.2 & 1 \\ 
  G018.152$+$00.090 & 1G018154$+$000992 &  & 18.154 & 0.099 & 53.0 & 4.1 & 4.6 & 1.32 & 1.6 & 0.200 & 47.45 & 0.2 & 0.4 & 4 \\ 
  G018.451$-$00.016 & 1G018452$-$000152 &  & 18.452 & $-$0.015 & 56.5 & 11.9 & 4.8 & 1.05 & 3.6 & 0.485 & 48.76 & 0.6 & 0.7 & 4 \\ 
  G018.657$-$00.057 & 1G018658$-$000495 &  & 18.658 & $-$0.050 & 44.1 & 12.5 & 5.3 & 0.52 & 1.9 & 0.146 & 48.28 & 0.4 & 0.3 & 1 \\ 
  G018.741$+$00.250 & 1G018743$+$002521 &  & 18.743 & 0.252 & 19.1 & 14.2 & 6.9 & 0.42 & 1.8 & 0.099 & 48.22 & 0.4 & 0.3 & 1 \\ 
  G019.504$-$00.193 & 1G019505$-$001900 & N25 & 19.505 & $-$0.190 & 37.8 & 12.8 & 5.7 & 0.79 & 2.9 & 0.110 & 48.18 & 0.4 & 0.7 & 1 \\ 
  G019.629$-$00.095 & 1G019632$-$001189 &  & 19.632 & $-$0.119 & 58.6 & 11.7 & 4.8 & 3.94 & 13.4 & 4.028 & 49.66 & 1.1 & 4.3 & 3 \\ 
  G020.227$+$00.110 & 1G020223$+$001118 &  & 20.223 & 0.112 & 22.1 & 13.9 & 6.8 & 0.84 & 3.4 & 0.083 & 48.13 & 0.3 & 0.9 & 4 \\ 
  G020.988$+$00.092 & 1G020991$+$000963 &  & 20.991 & 0.096 & 18.6 & 14.1 & 7.0 & 1.33 & 5.5 & 1.396 & 49.36 & 0.9 & 1.0 & 1 \\ 
  G022.761$-$00.492 & 1G022756$-$004827 &  & 22.756 & $-$0.483 & 74.8 & 4.8 & 4.3 & 1.99 & 2.8 & 2.769 & 48.73 & 0.6 & 0.4 & 5 \\ 
  G022.988$-$00.360 & 1G022991$-$003666 &  & 22.991 & $-$0.367 & 74.1 & 4.8 & 4.3 & 1.95 & 2.7 & 1.039 & 48.30 & 0.4 & 0.6 & 4 \\ 
  G023.661$-$00.252 & 1G023660$-$002527 &  & 23.660 & $-$0.253 & 66.2 & 11.2 & 4.9 & 0.90 & 2.9 & 0.224 & 48.37 & 0.4 & 0.6 & 1 \\ 
  G023.787$+$00.223 & 1G023798$+$002263 &  & 23.798 & 0.226 & 107.4 & 9.4 & 3.8 & 2.28 & 6.2 & 0.745 & 48.74 & 0.6 & 1.9 & 4 \\ 
  G024.728$+$00.159 & 1G024731$+$001580 &  & 24.731 & 0.158 & 109.3 & 6.3 & 3.7 & 0.78 & 1.4 & 0.110 & 47.56 & 0.2 & 0.3 & 5 \\ 
  G024.734$+$00.087 & 1G024735$+$000889 &  & 24.735 & 0.089 & 111.3 & 6.4 & 3.7 & 1.93 & 3.6 & 0.818 & 48.45 & 0.4 & 0.9 & 1 \\ 
  G025.397$+$00.033 & 1G025399$+$000360 &  & 25.399 & 0.036 & $-$14.4 & 17.3 & 10.4 & 1.62 & 8.2 & 1.097 & 49.44 & 0.9 & 2.0 & 3 \\ 
  G027.476$+$00.179 & 1G027484$+$001817 &  & 27.484 & 0.182 & 34.0 & 12.6 & 6.5 & 3.73 & 13.7 & 4.424 & 49.77 & 1.2 & 4.2 & 4 \\ 
  G027.682$+$00.076 & 1G027688$+$000778 &  & 27.688 & 0.078 & 100.2 & 6.0 & 4.1 & 0.96 & 1.7 & 0.119 & 47.55 & 0.2 & 0.4 & 5 \\ 
  G027.980$+$00.080 & 1G027981$+$000753 &  & 27.981 & 0.075 & 76.6 & 10.3 & 4.9 & 0.54 & 1.6 & 0.229 & 48.31 & 0.4 & 0.2 & 1 \\ 
  G027.997$+$00.317 & 1G028006$+$003153 &  & 28.006 & 0.315 & 92.7 & 9.4 & 4.4 & 2.03 & 5.5 & 0.369 & 48.43 & 0.4 & 1.9 & 4 \\ 
  G028.146$+$00.146 & 1G028140$+$001526 &  & 28.140 & 0.153 & 89.2 & 5.4 & 4.4 & 1.70 & 2.7 & 0.287 & 47.84 & 0.3 & 0.7 & 6 \\ 
  G028.638$+$00.194 & 1G028636$+$002033 &  & 28.636 & 0.203 & 103.8 & 7.5 & 4.0 & 1.80 & 3.9 & 0.412 & 48.29 & 0.4 & 1.1 & 4 \\ 
  G030.252$+$00.053 & 1G030250$+$000547 &  & 30.250 & 0.241 & 65.2 & 4.5 & 5.0 & 0.59 & 0.8 & 0.096 & 47.21 & 0.2 & 0.1 & 5 \\ 
  G030.311$-$00.215 & 1G030323$-$002072 &  & 30.329 & $-$0.209 & 106.1 & 7.3 & 4.2 & 0.91 & 1.9 & 0.153 & 47.83 & 0.3 & 0.4 & 4 \\ 
  G030.468$+$00.394 & 1G030461$+$004237 &  & 30.461 & 0.424 & 57.7 & 3.8 & 5.4 & 3.53 & 3.9 & 1.334 & 48.21 & 0.4 & 1.1 & 6* \\ 
  G030.690$-$00.258 & 1G030699$-$002560 &  & 30.699 & $-$0.256 & 98.5 & 8.4 & 4.4 & 2.17 & 5.3 & 2.713 & 49.20 & 0.8 & 1.1 & 4 \\ 
  G031.138$+$00.285 & 1G031147$+$002879 & N54 & 31.147 & 0.288 & 104.7 & 7.3 & 4.3 & 1.89 & 4.0 & 1.192 & 48.72 & 0.5 & 0.9 & 3 \\ 
  G031.470$-$00.344 & 1G031473$-$003459 &  & 31.473 & $-$0.346 & 88.8 & 9.0 & 4.7 & 1.12 & 2.9 & 0.477 & 48.51 & 0.5 & 0.6 & 4 \\ 
  G032.057$+$00.077 & 1G032057$+$000783 &  & 32.057 & 0.078 & 96.3 & 7.2 & 4.4 & 0.94 & 2.0 & 0.263 & 48.06 & 0.3 & 0.4 & 5 \\ 
  G032.152$+$00.131 & 1G032158$+$001306 &  & 32.158 & 0.131 & 95.0 & 6.1 & 4.5 & 0.78 & 1.4 & 0.740 & 48.36 & 0.4 & 0.2 & 5 \\ 
  G032.582$+$00.001 & 1G032584$+$000057 & N56 & 32.584 & 0.006 & 77.4 & 9.4 & 5.1 & 1.09 & 3.0 & 0.163 & 48.08 & 0.3 & 0.8 & 6 \\ 
  G032.733$+$00.209 & 1G032731$+$002120 &  & 32.731 & 0.212 & 16.1 & 13.2 & 7.7 & 0.45 & 1.7 & 0.030 & 47.64 & 0.2 & 0.4 & 4 \\ 
  G032.928$+$00.606 & 1G032929$+$006055 &  & 32.929 & 0.605 & $-$38.3 & 19.2 & 13.0 & 0.45 & 2.5 & 0.293 & 48.95 & 0.7 & 0.3 & 5 \\ 
  G033.813$-$00.150 & 1G033815$-$001494 & N60 & 33.815 & $-$0.149 & 50.0 & 10.8 & 6.0 & 0.41 & 1.3 & 0.040 & 47.59 & 0.2 & 0.2 & 5 \\ 
  G034.089$+$00.438 & 1G034088$+$004405 &  & 34.088 & 0.441 & 32.6 & 11.8 & 6.8 & 0.69 & 2.4 & 0.219 & 48.41 & 0.4 & 0.4 & 3 \\ 
  G034.133$+$00.471 & 1G034132$+$004724 &  & 34.132 & 0.472 & 34.6 & 11.7 & 6.7 & 0.40 & 1.4 & 0.586 & 48.83 & 0.6 & 0.1 & 3 \\ 
  G034.256$+$00.136 & 1G034262$+$001267 &  & 34.261 & 0.136 & 54.0 & 3.5 & 5.8 & 2.08 & 2.1 & 10.758 & 49.04 & 0.7 & 0.2 & 4 \\ 
  G035.051$-$00.520 & 1G035051$-$005195 &  & 35.051 & $-$0.519 & 48.0 & 10.7 & 6.2 & 0.39 & 1.2 & 0.215 & 48.31 & 0.4 & 0.1 & 5 \\ 
  G035.467$+$00.004 & 1G035476$+$000027 &  & 35.476 & 0.003 & 58.6 & 10.0 & 5.8 & 1.08 & 3.1 & 0.169 & 48.15 & 0.4 & 0.8 & 4 \\ 
  G035.543$+$00.006 & 1G035544$+$000131 & N67 & 35.544 & 0.013 & 57.5 & 10.1 & 5.9 & 0.87 & 2.5 & 0.218 & 48.27 & 0.4 & 0.5 & 3 \\ 
  G035.571$+$00.071 & 1G035573$+$000703 &  & 35.573 & 0.070 & 47.6 & 10.7 & 6.2 & 0.48 & 1.5 & 0.455 & 48.64 & 0.5 & 0.1 & 5 \\ 
  G035.649$-$00.053 & 1G035652$-$000348 & N68 & 35.652 & $-$0.035 & 51.9 & 10.4 & 6.1 & 4.21 & 12.7 & 2.853 & 49.41 & 0.9 & 4.6 & 6 \\ 
  G037.259$-$00.083 & 1G037261$-$000809 &  & 37.261 & $-$0.081 & 40.8 & 10.8 & 6.5 & 0.81 & 2.6 & 0.122 & 48.07 & 0.3 & 0.6 & 4 \\ 
  G037.344$+$00.684 & 1G037349$+$006876 &  & 37.349 & 0.688 & 45.6 & 10.5 & 6.4 & 0.60 & 1.8 & 0.176 & 48.21 & 0.4 & 0.3 & 5 \\ 
  G037.750$-$00.110 & 1G037751$-$001098 & N70 & 37.751 & $-$0.110 & 49.7 & 10.1 & 6.2 & 0.87 & 2.6 & 0.374 & 48.50 & 0.5 & 0.4 & 3 \\ 
  G037.754$+$00.560 & 1G037754$+$005577 &  & 37.755 & 0.560 & 18.3 & 12.2 & 7.6 & 0.38 & 1.4 & 0.054 & 47.82 & 0.3 & 0.2 & 4 \\ 
  G038.045$-$00.034 & 1G038046$-$000318 &  & 38.046 & $-$0.032 & 58.3 & 9.5 & 5.9 & 1.57 & 4.3 & 0.268 & 48.31 & 0.4 & 1.3 & 4 \\ 
  G038.840$+$00.495 & 1G038839$+$004990 &  & 38.839 & 0.499 & $-$43.2 & 18.2 & 12.8 & 0.58 & 3.1 & 0.041 & 48.05 & 0.3 & 0.8 & 3 \\ 
  G039.728$-$00.396 & 1G039728$-$003965 &  & 39.728 & $-$0.397 & 58.3 & 9.1 & 6.0 & 0.48 & 1.3 & 0.253 & 48.24 & 0.4 & 0.1 & 5 \\ 
  G039.864$+$00.645 & 1G039868$+$006467 &  & 39.868 & 0.647 & $-$40.9 & 17.6 & 12.4 & 0.46 & 2.4 & 0.128 & 48.52 & 0.5 & 0.4 & 5 \\ 
  G039.873$-$00.177 & 1G039874$-$001737 &  & 39.874 & $-$0.174 & 60.0 & 9.0 & 5.9 & 0.82 & 2.1 & 0.110 & 47.87 & 0.3 & 0.5 & 4 \\ 
  G041.235$+$00.367 & 1G041237$+$003647 &  & 41.237 & 0.365 & 71.3 & 5.4 & 5.5 & 1.84 & 2.9 & 0.375 & 47.96 & 0.3 & 0.8 & 6 \\ 
  G041.512$+$00.021 & 1G041519$+$000375 & N79 & 41.519 & 0.037 & 17.7 & 11.6 & 7.7 & 2.06 & 7.0 & 0.602 & 48.83 & 0.6 & 2.2 & 6 \\ 
  G041.929$+$00.030 & 1G041931$+$000351 & N80 & 41.931 & 0.035 & 20.7 & 11.3 & 7.6 & 1.45 & 4.8 & 0.193 & 48.31 & 0.4 & 1.5 & 6 \\ 
  G042.103$-$00.623 & 1G042104$-$006230 & N82 & 42.104 & $-$0.623 & 66.0 & 7.7 & 5.8 & 1.81 & 4.1 & 1.389 & 48.84 & 0.6 & 0.8 & 3 \\ 
  G042.111$-$00.444 & 1G042112$-$004451 & N83 & 42.112 & $-$0.445 & 53.4 & 8.9 & 6.2 & 0.49 & 1.3 & 0.181 & 48.08 & 0.3 & 0.2 & 5 \\ 
  G042.434$-$00.275 & 1G042426$-$002691 &  & 42.426 & $-$0.269 & 61.5 & 8.1 & 5.9 & 1.23 & 2.9 & 0.715 & 48.59 & 0.5 & 0.5 & 6 \\ 
  G043.185$-$00.525 & 1G043185$-$005229 &  & 43.185 & $-$0.523 & 59.1 & 8.1 & 6.0 & 0.55 & 1.3 & 0.713 & 48.59 & 0.5 & 0.1 & 5 \\ 
  G043.432$+$00.516 & 1G043434$+$005189 &  & 43.434 & 0.519 & $-$12.8 & 13.6 & 9.5 & 1.06 & 4.2 & 0.104 & 48.21 & 0.4 & 1.3 & 6 \\ 
  G043.774$+$00.057 & 1G043775$+$000606 & N90 & 43.775 & 0.061 & 70.5 & 6.1 & 5.7 & 1.37 & 2.4 & 0.153 & 47.68 & 0.2 & 0.7 & 6* \\ 
  G043.792$-$00.122 & 1G043797$-$001171 &  & 43.789 & $-$0.116 & 43.3 & 9.2 & 6.6 & 0.56 & 1.5 & 0.064 & 47.65 & 0.2 & 0.3 & 5 \\ 
  G043.818$+$00.395 & 1G043829$+$003957 &  & 43.829 & 0.396 & $-$10.5 & 13.3 & 9.3 & 1.26 & 4.9 & 0.216 & 48.50 & 0.5 & 1.4 & 6 \\ 
  G044.418$+$00.535 & 1G044417$+$005356 &  & 44.417 & 0.536 & $-$55.1 & 18.4 & 13.8 & 0.47 & 2.5 & 0.042 & 48.08 & 0.3 & 0.6 & 3 \\ 
  G044.501$+$00.332 & 1G044503$+$003373 &  & 44.503 & 0.337 & $-$43.0 & 16.6 & 12.2 & 1.29 & 6.2 & 0.246 & 48.75 & 0.6 & 1.9 & 6 \\ 
  G045.118$+$00.144 & 1G045123$+$001453 &  & 45.123 & 0.145 & 56.7 & 7.7 & 6.2 & 1.19 & 2.7 & 3.881 & 49.28 & 0.8 & 0.3 & 3 \\ 
  G045.197$+$00.740 & 1G045204$+$007443 &  & 45.204 & 0.744 & $-$35.0 & 15.5 & 11.3 & 1.17 & 5.3 & 0.177 & 48.55 & 0.5 & 1.6 & 4 \\ 
  G045.391$-$00.725 & 1G045387$-$007155 & N95 & 45.387 & $-$0.715 & 52.5 & 8.0 & 6.3 & 1.69 & 3.9 & 0.276 & 48.17 & 0.4 & 1.2 & 6 \\ 
  G045.475$+$00.130 & 1G045476$+$001340 &  & 45.476 & 0.134 & 55.6 & 7.7 & 6.2 & 1.00 & 2.2 & 1.747 & 48.94 & 0.6 & 0.3 & 3 \\ 
  G045.825$-$00.291 & 1G045822$-$002892 &  & 45.822 & $-$0.289 & 61.2 & 6.6 & 6.0 & 0.71 & 1.4 & 0.336 & 48.09 & 0.3 & 0.2 & 3 \\ 
  G045.933$-$00.403 & 1G045934$-$004012 &  & 45.934 & $-$0.401 & 63.9 & 5.9 & 6.0 & 0.68 & 1.2 & 0.071 & 47.32 & 0.2 & 0.2 & 5 \\ 
  G046.495$-$00.241 & 1G046481$-$002389 &  & 46.481 & $-$0.239 & 57.2 & 4.7 & 6.1 & 3.66 & 5.0 & 2.094 & 48.59 & 0.5 & 1.4 & 6 \\ 
  
\end{longtable}
%\end{landscape}

%\clearpage
%\newpage

\section{Region Images}\label{app:group_images}

Images of the 76 \hii regions may be found in the online version of this paper.

%\input{appendiximages}

%%%%%%%%%%%%%%%%%%%%%%%%%%%%%%%%%%%%%%%%%%%%%%%%%%

% Don't change these lines
\bsp	% typesetting comment
\label{lastpage}
\end{document}